\documentclass[11pt,a4paper]{article}
\usepackage{jheppub}
\usepackage{graphicx, bm, color, amssymb, amsmath,slashed,xcolor}
\usepackage{epstopdf}
\usepackage{multirow}
\usepackage{float}
\usepackage[caption=false]{subfig}
\usepackage{tabularx}
\usepackage{hyperref}
\usepackage[normalem]{ulem}
\newcommand{\lsim}{\mbox{\raisebox{-.6ex}{~$\stackrel{<}{\sim}$~}}}

\newcolumntype{L}[1]{>{\raggedright\arraybackslash}p{#1}}
\newcolumntype{C}[1]{>{\centering\arraybackslash}p{#1}}
\newcolumntype{R}[1]{>{\raggedleft\arraybackslash}p{#1}}

\begin{document}
\preprint{IP/BBSR/2019-21  \newline
 \hspace*{11.9cm} }

\title{Same-Sign Tetralepton Signature at Large Hadron Collider, and future $pp$ Collider. }
\author[a]{Eung Jin Chun,}
\author[b,c]{Sarif Khan,}
\author[d,e]{Sanjoy Mandal,}
\author[d,e]{Manimala Mitra,}
\author[d,e]{Sujay Shil}

\affiliation[a]{Korea Institute for Advanced Study, Seoul 130-722, Korea}
\affiliation[b]{Institut f\"{u}r Theoretische Physik,
Georg-August-Universit\"{a}t G\"{o}ttingen, Friedrich-Hund-Platz 1,
G\"{o}ttingen, D-37077 Germany}
\affiliation[c]{Harish-Chandra Research Institute, Allahabad, India}
\affiliation[d]{Institute of Physics, Sachivalaya Marg, Bhubaneswar, Odisha 751005, India}
\affiliation[e]{Homi Bhabha National Institute, BARC Training School Complex, Anushakti Nagar, Mumbai 400094, India}

\emailAdd{ejchun@kias.re.kr}
\emailAdd{sarif.khan@uni-goettingen.de}
\emailAdd{smandal@iopb.res.in}
\emailAdd{manimala@iopb.res.in}
\emailAdd{sujayshil1@gmail.com}

\abstract{We analyze a novel signature of the type II seesaw model - same-sign tetra-lepton signal arising from the mixing of neutral Higgs bosons and their subsequent decays to singly and doubly charged Higgs bosons. For this, we consider wide ranges of the triplet vacuum expectation value (vev)
and Yukawa couplings, that are consistent with the observed neutrino masses and mixing as well as the LHC search limits. 
We find that a doubly charged Higgs boson with mass around 250 GeV and triplet vev around $10^{-4}-10^{-2}$ GeV can give significantly large number of events through it decay to same-sign $W$ gauge bosons at High-Luminosity LHC with $3000 \text{fb}^{-1}$ of data. We also pursue the analysis for a future hadron collider with the c.m. energy of 100 TeV. 
Considering a heavy Higgs boson around 900 GeV and an intermediate region of the triplet vev, where both same-sign dilepton and gauge boson decays can occur, we identify a limited range of the parameters where the number of same-sign tetra-lepton events are as large as 1000.} 

\maketitle
\section{Introduction \label{intro}}

After the discovery of the Standard Model (SM) Higgs boson,  one of the key questions that still remains unexplained   is the origin of light neutrino masses and mixings.  A number of neutrino oscillation experiments have  observed that the  solar and atmospheric neutrino mass splittings are $\Delta m^2_{12} \sim 10^{-5}$ $ \rm{eV}^2$ and $\Delta m^2_{13} \sim 10^{-3}$ $ \rm{eV}^2$. The   PMNS mixing angles are $\theta_{12} \sim 32^\circ$, $\theta_{23} \sim 45^\circ$, and $\theta_{13} \sim 9^\circ$ \cite{deSalas:2017kay}.  Once we include the right handed neutrinos in the theory, a Dirac mass term can be generated for light neutrinos.  However, to generate eV neutrino masses, this
requires a very 
 large hierarchy of the Yukawa couplings  $Y_{\nu} \sim \mathcal{O}(10^{-11})$ within the SM.  The light neutrinos, being electromagnetic charge neutral, can be Majorana particle, and  their masses can  have a different origin compared to the other  SM fermions. One of the   profound mechanisms to generate Majorana masses of the light neutrinos is seesaw, where 
 tiny eV masses of the Majorana neutrinos  are generated from  lepton number violating (LNV) $d=5$ operator $LLHH/\Lambda$~\cite{Weinberg:1979sa,Wilczek:1979hc}.   There can be different UV completed 
 theories behind this operator, commonly known as, type-I, -II, and -III seesaw mechanisms. These different models accommodate  extensions of the SM fermion/scalar 
 contents by   $SU(2)_L$
singlet fermions~\cite{Minkowski:1977sc,Mohapatra:1979ia,Yanagida:1979as,GellMann:1980vs,Schechter:1980gr,Babu:1993qv,Antusch:2001vn}, $SU(2)_L$ triplet scalar boson~\cite{Magg:1980ut,Cheng:1980qt,Lazarides:1980nt,Mohapatra:1980yp}, and $SU(2)_L$ triplet fermions~\cite{Foot:1988aq}, respectively. Among these,  type-II seesaw model, where a triplet scalar field with the hypercharge $Y = +2$ is added to the SM, has an  extended scalar sector. See \cite{Arhrib:2011uy,Dev:2013ff,Das:2016bir} for the details of the Higgs spectrum. The bound from vacuum stability, perturbativity, and electroweak precision test has been studied in~\cite{Chun:2012jw}. The neutral component of the triplet acquires a vacuum expectation value (vev) $v_{\Delta}$, and generates eV scale neutrino masses through the Yukawa interaction between  lepton doublets and triplet Higgs field. The same Yukawa interaction  also have a large impact on  the  charged Higgs phenomenology in this model. The presence of a doubly charged Higgs,  that can have distinct decay modes whose branching ratios are determined by the observed neutrino oscillation data~\cite{Chun:2003ej}, is the most appealing feature of this model. Hence a  discovery of this exotic particle will be a smoking gun signature  of this model. 
 
A number of searches  have  been carried out  to look  for  the signature of the doubly charged Higgs at collider and non-collider experiments~\cite{Chun:2003ej}. See   \cite{Akeroyd:2005gt} for Tevatron,  and  \cite{Perez:2008ha,Melfo:2011nx,delAguila:2008cj,Chakrabarti:1998qy,Aoki:2011pz,Akeroyd:2011zza,Chun:2013vma,delAguila:2013mia, Banerjee:2013hxa,kang:2014jia,Han:2015hba,Han:2015sca,Babu:2016rcr} for LHC, \cite{Crivellin:2018ahj, Borah:2018yxd,  Padhan:2019jlc} for HE-LHC and future hadron colliders.  Depending on the triplet vev, the doubly-charged Higgs boson can decay via  distinguished   decays modes. Assuming degenerate charged Higgs masses, it decays pre-dominantly to same-sign dileptons~(gauge bosons) for $v_{\Delta}<(>)\,10^{-4}$~GeV. For non-degenerate charged Higgs, in the intermediate range of triplet vev, the  cascade decay to singly charged Higgs can also be dominant and {has been explored in}  \cite{Perez:2008ha,Melfo:2011nx,delAguila:2008cj}. The CMS and ATLAS collaborations have searched for the same-sign dilepton final states, and constrained the mass of the doubly-charged Higgs as $M_{H^{\pm \pm}} > 820,  870\, \rm{GeV} $  at  95$\%$ C.L. \cite{Aaboud:2017qph,CMS-PAS-HIG-16-036} assuming $\text{Br}(H^{\pm\pm}\to\ell^{\pm}\ell^{\pm})=100\%$. The   vector boson fusion channel, where the  $H^{\pm \pm}$ is produced in association with two jets,  gives relaxed constraints \cite{Khachatryan:2014sta, Sirunyan:2017ret}.  The collider signatures and the discovery prospect of this scenario have been discussed in \cite{Kanemura:2013vxa,Kanemura:2014goa,Kanemura:2014ipa}, and \cite{Mitra:2016wpr, Ghosh:2017pxl}. Previous searches for $H^{\pm \pm}$ in the pair-production channel and their subsequent decays into same-sign leptons at LEP-II has put a constraint $M_{H^{\pm \pm}} > 97.3 $ GeV at $95 \%$ C.L. \cite{Abdallah:2002qj}. {For the earlier discussions on Higgs triplet model at a linear collider, see \cite{Shen:2015bna,Blunier:2016peh,Cao:2016hvg,Guo:2016hjt,Agrawal:2018pci}.

Most of the works in the literature explored di-lepton or gauge boson decay modes of the doubly charged Higgs, leading to multi-lepton final states. Due to the possible cascade decays of the charge neutral  Higgs into a  singly charged Higgs, and the cascade decay of a singly charged Higgs into a doubly charged  Higgs, the model can also lead to a very unique signature, same-sign tetra-lepton final states. This has been first proposed in \cite{Chun:2012zu}, and explored for the lower triplet vev, where di-lepton decay is pre-dominant.  In this work, we consider a wide range of triplet vev, in particularly, focussing on gauge boson decay modes, and explore the signature for 14 TeV LHC. For higher range of triplet vev, as the  LHC constraint on the mass of doubly charged Higgs is relatively relaxed, we 
therefore perform the analysis for lighter Higgs state, as low as $M_{H^{\pm \pm}} \sim {247} $ GeV. In addition, we also consider a very high energy $pp$ collider, that can operate with c.m.energy $\sqrt{s}=100$ TeV, and explore this unique signature for a heavy doubly charged Higgs. We show that for heavier doubly charged Higgs, there is a very narrow region of triplet vev,  which can accommodate significantly large $\mathcal{O}(10^3)$ same-sign tetra-lepton signatures. 

Our paper is organized as follows: we briefly review the basics of the type-II seesaw model in Sec.~\ref{model}. In Sec.~\ref{expcons}, we discuss branching ratios of doubly and singly charged Higgs, and the   relation between $H^{\pm\pm}$ and $H^{\pm}$ decays. In Sec.~\ref{same sign tetralepton section}, and in Sec.~\ref{signal at 100 TeV},  we present the simulation of same-sign tetra-lepton  signal at $\sqrt{s}=14$ TeV LHC, and $\sqrt{s}=100$ TeV. Finally, we present our conclusions in Sec.~\ref{conclu}.

\section{Model Description {\label{model}}}
 One of the most simplest seesaw models is the type-II seesaw model \cite{Magg:1980ut,Cheng:1980qt,Lazarides:1980nt,Mohapatra:1980yp}, that, in addition to the SM particle contents, also  contains one   $SU(2)_L$ triplet Higgs field
\begin{eqnarray}
 \Delta=\begin{pmatrix} \frac{\Delta^+}{\sqrt{2}} & \Delta^{++} \\ \Delta^0 & -\frac{\Delta^+}{\sqrt{2}}
\end{pmatrix}  ~~~ \sim (1,3,2).
\end{eqnarray}
 The neutral components of the SM doublet ($\Phi$) and triplet Higgs fields are denoted as   $\Phi^0=\frac{1}{\sqrt{2}}(\phi^0+i\chi^0)$ and $\Delta^0=\frac{1}{\sqrt{2}}(\delta^0+i\eta^0)$, respectively. The neutral component of $\Delta$ acquires vev and generates Majorana masses for light neutrinos.  We denote the vevs of  $\phi^0$ and $\delta^0$ by  $v_{\Phi}$ and $v_{\Delta}$, where   $v^2=v^2_{\Phi}+v^2_{\Delta}=(246 \, \, \rm{GeV})^2$. The  kinetic term 
 for the triplet has the following form
\begin{eqnarray}
\mathcal{L}_{\rm{kin}}( \Delta)&=&\rm{Tr}[(D_\mu \Delta)^\dagger (D^\mu \Delta)], 
\label{kinetic}
\end{eqnarray}
In the above, $D_{\mu}$ is the co-variant derivative  $D_\mu \Delta=\partial_\mu \Delta+i\frac{g}{2}[\tau^aW_\mu^a,\Delta]+ig'B_\mu\Delta$. The new triplet scalar field $\Delta$, being a triplet under $SU(2)_L$ interacts  with the 
SM gauge bosons. In addition to the kinetic term, $\Delta$  has  Yukawa interaction with the SM lepton doublet. 
The Yukawa interactions of $\Delta$ with the lepton fields are 
\begin{eqnarray}
\mathcal{L}_Y(\Phi, \Delta)&=& Y_{\Delta}\overline{L_L^{c}}i\tau_2\Delta L_L+\rm{h.c.},~~~~ 
\label{yukawa}
\end{eqnarray}
where  $Y_{\Delta}$ is a $3\times 3$ matrix and $c$ denotes charge conjugation.  The scalar potential of the Higgs fields $\Phi$ and $\Delta$ is  
\begin{eqnarray}
V(\Phi,\Delta)&=&m_\Phi^2\Phi^\dagger\Phi+\tilde{M}^2_{\Delta}\rm{Tr}(\Delta^\dagger\Delta)+\left(\mu \Phi^Ti\tau_2\Delta^\dagger \Phi+\rm{h.c.}\right)+\frac{\lambda}{4}(\Phi^\dagger\Phi)^2 \nonumber\\
&+&\lambda_1(\Phi^\dagger\Phi)\rm{Tr}(\Delta^\dagger\Delta)+\lambda_2\left[\rm{Tr}(\Delta^\dagger\Delta)\right]^2 +\lambda_3\rm{Tr}[(\Delta^\dagger\Delta)^2]
+\lambda_4\Phi^\dagger\Delta\Delta^\dagger\Phi,~~~~
\label{eqn:scalpt}
\end{eqnarray}
where $m_{\Phi}$ and $\tilde{M}_{\Delta}$ are real parameters with mass dimension {1},  and $\lambda$, $\lambda_{1-4}$ are dimensionless quartic Higgs couplings.  Note that, $\mu$ is the parameter with positive mass dimension. The triplet field $\Delta$ carries lepton number +2 and hence the Yukawa term conserves lepton number. However, the lepton number is violated  2-units by a non-zero  $\mu$.  Therefore, together a non-zero $\mu$ and a non-zero 
$Y_{\nu}$ violate lepton number symmetry. 

The scalar potential that generates scalar mass matrix, includes  tri-linear as well as quartic couplings among the scalar fields. The scalar mass matrix, after diagonalization, generates 
 seven physical Higgs states.  They are: the charged Higgs bosons $H^{\pm \pm}$, $H^{\pm}$, and the neutral Higgs bosons $h^0, H^0 $ and $A^0$. The two charged scalar fields
$\Phi^{\pm}$ of $\Phi$ and $\Delta^{\pm}$ of $\Delta$  mix to give singly-charged states $H^{\pm}$  and the charged Goldstone $\chi^{\pm}$ bosons. Similarly, the mixing between the two CP-odd fields ($\chi^{0}$ and $\eta^{0}$)
gives rise to  $A^0$, and the neutral Goldstone boson $\rho^0$. Finally, we obtain the SM Higgs boson ($h$) and a heavy Higgs boson ($H$) via the mixing of the two neutral CP-even states $\Phi^{0}$ and $\delta^{0}$.  For 
the detail description of the charged and neutral mass matrix, see \cite{Arhrib:2011uy}.

 The minimization  conditions of the potential are   $$\frac {\partial V(\Phi,\Delta)}{\partial v_\Phi}=0, \qquad \frac {\partial V(\Phi,\Delta)}{\partial v_\Delta}=0.$$ These give the following conditions  for $m_\Phi^2, M^2$:
\begin{eqnarray}
m_\Phi^2&=&\displaystyle \frac{1}{2}\left[-\frac{v_\Phi^2\lambda}{2}-v_\Delta^2(\lambda_1+\lambda_4)+2\sqrt{2}\mu v_\Delta\right],\\
\tilde{M}^2&=&\displaystyle M_\Delta^2-\frac{1}{2}\left[2v_\Delta^2(\lambda_2+\lambda_3)+v_\Phi^2(\lambda_1+\lambda_4)\right], \label{vc}
\rm{ with }~ M_\Delta^2\equiv \frac{v_\Phi^2\mu}{\sqrt{2}v_\Delta}. 
\end{eqnarray}
 The diagonalization conditions for the neutral and charged scalar fields are,
\begin{eqnarray}
\left(
\begin{array}{c}
\phi^\pm\\
\Delta^\pm
\end{array}\right)&=&
\left(
\begin{array}{cc}
\cos \beta_\pm & -\sin\beta_\pm \\
\sin\beta_\pm   & \cos\beta_\pm
\end{array}
\right)
\left(
\begin{array}{c}
\chi^\pm\\
H^\pm
\end{array}\right),\quad 
\left(
\begin{array}{c}
\chi\\
\eta
\end{array}\right)=
\left(
\begin{array}{cc}
\cos \beta_0 & -\sin\beta_0 \\
\sin\beta_0   & \cos\beta_0
\end{array}
\right)
\left(
\begin{array}{c}
\rho^0\\
A^0
\end{array}\right),\nonumber\\
\left(
\begin{array}{c}
\phi^0\\
\delta^0
\end{array}\right)&=&
\left(
\begin{array}{cc}
\cos \alpha & -\sin\alpha \\
\sin\alpha   & \cos\alpha
\end{array}
\right)
\left(
\begin{array}{c}
h^0\\
H^0
\end{array}\right),
\end{eqnarray}
where   the mixing angles
\begin{eqnarray}
\tan\beta_\pm=\frac{\sqrt{2}v_\Delta}{v_\Phi},\quad \tan\beta_0 = \frac{2v_\Delta}{v_\Phi}, \quad
\tan2\alpha &=&\frac{4v_\Delta}{v_\Phi}\frac{v_\Phi^2(\lambda_1+\lambda_4)-2M_\Delta^2}{v_\Phi^2\lambda-2M_\Delta^2-4v_\Delta^2(\lambda_2+\lambda_3)}.~~~~~~~~ \label{tan2a}
\label{tan-beta-alpha}
\end{eqnarray}
All these  mixings being proportionl to the ratio of $\frac{v_{\Delta}}{v_{\Phi}}$ is very small. 

The physical masses of {the} doubly and singly charged Higgs bosons $H^{\pm \pm}$ and $H^{\pm}$  can be written as
\begin{eqnarray}
m_{H^{++}}^2=M_\Delta^2-v_\Delta^2\lambda_3-\frac{\lambda_4}{2}v_\Phi^2,\label{eq:mhpp}~~~
m_{H^+}^2= \left(M_\Delta^2-\frac{\lambda_4}{4}v_\Phi^2\right)\left(1+\frac{2v_\Delta^2}{v_\Phi^2}\right).\label{eq:mhp}
\end{eqnarray}
The CP-even and CP-odd neutral Higgs bosons $h$, and  $H$  have the physical masses
\begin{eqnarray}
m_{h^0}^2=\mathcal{T}_{11}^2\cos^2\alpha+\mathcal{T}_{22}^2\sin^2\alpha-\mathcal{T}_{12}^2\sin2\alpha, \label{mh}~\\
m_{H^0}^2=\mathcal{T}_{11}^2\sin^2\alpha+\mathcal{T}_{22}^2\cos^2\alpha+\mathcal{T}_{12}^2\sin2\alpha.\label{mH}
\end{eqnarray}
In the above $\mathcal{T}_{11}$,  $\mathcal{T}_{22}$ and $\mathcal{T}_{12}$ have the following expressions: 
 \begin{eqnarray}
\mathcal{T}_{11}^2=\frac{v_\Phi^2\lambda}{2},~~
\mathcal{T}_{22}^2=M_\Delta^2+2v_\Delta^2(\lambda_2+\lambda_3), ~~
\mathcal{T}_{12}^2=-\frac{2v_\Delta}{v_\Phi}M_\Delta^2+v_\Phi v_\Delta(\lambda_1+\lambda_4).
\end{eqnarray}
The CP-odd Higgs field $A^0$ has the following mass
\begin{eqnarray}
m_A^2 &= &M_\Delta^2\left(1+\frac{4v_\Delta^2}{v_\Phi^2}\right) \label{mA},\quad \mathrm{with}~~ M^2_{\Delta}=\frac{v^2_{\Phi} \mu}{\sqrt{2} v_{\Delta}}.
\end{eqnarray}

{The  difference between $H^{\pm \pm}$ and $H^{\pm}$ masses  is dictated by the coupling $\lambda_4$ of the scalar potential.  For a positive $\lambda_4$, the ${H^{\pm \pm}}$ is lighter than ${H^{\pm }}
$.} The mass difference $\Delta M^2$ is 
\begin{equation}
\Delta M^2=M^2_{H^{\pm}}-M^2_{H^{\pm \pm}} \sim \frac{\lambda_4}{2} v^2_{\Phi}+\mathcal{O}(v^2_{\Delta}).
\label{diffchdmass}
\end{equation}
Throughout our analysis, we consider the mass hierarchy $M_{H^{\pm \pm}} < M_{H^{\pm}}$.   Among the neutral Higgs fields, we identify $h^0$  as the SM Higgs with mass $M_{h^0} = 125$ GeV. The mass of $h^0$  is primarily decided by  $\lambda$, where the mass of $H^0$   is primarily decided by $M_{\Delta}$.  The neutral Higgs mixing angle $\alpha$ is very small, and hence, $\cos \alpha \simeq 1$.  On the other hand, the charged Higgs and CP odd Higgs mixing angles $\tan \beta_{\pm}$ and $\tan \beta_{0}$ being proportional $v_{\Delta}/v_{\Phi}$, is very small, $\tan \beta \sim 10^{-3}$. 
Note that, the mass square  difference between $H^{\pm}$ and $H^0$  in the limit $v_{\Delta} < v_{\Phi}$ is 
\begin{eqnarray} 
M^2_{H^{0}}-M^2_{H^{\pm}} \sim  \lambda_4 \frac{v^2_{\Phi}}{4}+\mathcal{O}({v^2_{\Delta}})
\end{eqnarray}
Therefore, the mass difference between $M_{H^{\pm \pm}}$, $M_{H^{\pm}}$ and the mass difference between $M_{H^{0}}$, $M_{H^{\pm}}$ are almost similar, and dictated by the same set of parameters $\lambda_4$, and electroweak vev
$v_{\Phi}$.  The mass square difference betwteen $H^0$ and $A^0$ is extremely small, {as this is proportional to the triplet vev}, 

\begin{eqnarray}
M^2_{H^0}- M^2_{A^0} \sim 2 v^2_{\Delta} (\lambda_2+\lambda_3) - \frac{4}{\sqrt{2}} \mu v_{\Delta}.
\end{eqnarray}

We denote the mass difference between $H^0$ and $A^0$ by $M_{H^{0}}-M_{A^0} \sim \delta M \sim v_{\Delta}$, and the mass difference between $H^{\pm} $ and $H^0$ by $M_{H^{\pm}}-M_{H^0} \sim \Delta M $.  As we will discuss in the next subsequent sections, the later parameter is important for few of the decay modes that depend on charged Higgs and neutral Higgs mass splitting, and is  one of the key parameter for our discussion.

Due to the non-trivial representations of $\Delta$, the Higgs triplet has interactions with a number of SM fermions and gauge bosons. This opens up a number of possible decay modes that can be explored at 
the LHC, and at other future  colliders. In the next section, we summarise the different direct experimental constraints on the  charged Higgs states. 

\section{Decay Modes and Experimental Constraints \label{expcons}}

We assume the neutral Higgs $H^0$ and $A^0$ are more massive than the charged Higgs. Among the charged Higgs,  $H^{\pm}$ is heavier than $H^{\pm \pm}$. 
 The doubly-charged Higgs boson $H^{\pm \pm }$ of this model  can decay into the leptonic or bosonic states and gives unique signatures at high energy colliders. 
The partial decay widths and  branching ratios of the $H^{\pm \pm }$ depend on the triplet vev $v_{\Delta}$.  For smaller triplet vev, the $H^{\pm \pm}$ predominantly decays into the same-sign leptonic states $H^{\pm \pm } \to l^{\pm} l^{\pm}$, whereas for  larger $v_{\Delta}$, the gauge boson mode $H^{\pm \pm} \to W^{\pm} W^{\pm}$ becomes dominant \cite{Chun:2003ej, Perez:2008ha, Melfo:2011nx}.  The relevant decay widths are calculated to be, 

\begin{equation}
\Gamma (H^{\pm \pm} \to l^{\pm}_i l^{\pm} _j)=\Gamma_{l_i l _j}=\frac{M_H^{\pm \pm} } {(1+\delta_{ij}) 8 \pi}   \left |\frac{M_{\nu_{ij}}}{v_{\Delta}} \right |^2, \, \, M_{\nu}=Y_{\Delta} v_{\Delta},
\end{equation}

\begin{equation}
\Gamma (H^{\pm \pm} \to W^{\pm} W^{\pm})=\Gamma_{W^{\pm }W^{\pm }}=\frac{g^2 v^2_{\Delta}}{8 \pi M_{H^{\pm \pm}}} \sqrt{1- \frac{4}{r^2_W}} \left[ \left (2+(r_W/2-1)^2 \right ) \right ].
\end{equation}
Here,  $M_{\nu}$ denotes the neutrino mass matrix, $i,j$ are the generation indices,  $\Gamma_{l_i l_j}$ and $\Gamma_{W^{\pm} W^{\pm}}$ are the partial decay widths for the $H^{\pm \pm} \to l^{\pm}_i l^{\pm}_j$, and $H^{\pm \pm} \to W^{\pm} W^{\pm}$  channels, respectively. The parameter  $r_W$ represents   the ratio of $H^{\pm \pm}$ and the $W$ gauge boson masses, $r_W=\frac{M_{H^{\pm \pm}}}{M_W}$.  

Other than the doubly charged Higgs, the model also contains a singly charged Higgs. The singly charged Higgs $H^{\pm}$ can decay to $l \nu$, $WZ, Wh, t \bar{b}$ final states.  Additionally, for non-degenerate charged Higgs masses, and triplet vev $v_{\Delta}$ in between $10^{-6} $ GeV and $10^{-2}$ GeV, the cascade decay $H^{\pm}\to H^{\pm \pm} W^*$ can also become  dominant.  The partial width for the charged  Higgs  decaying into $H^{\pm \pm} {W^{-}}^*$ have these following form:  

\begin{eqnarray}
\Gamma(H^{\pm} \to H^{\pm \pm} {W^{-}}^*)= \frac{9 g^4 M_{H^{\pm}}}{128 \pi^3} \cos^2 \beta_{\pm} G(\frac{M^2_{H^{\pm \pm}}}{M^2_{H^{\pm}}}, \frac{M^2_W}{M^2_{H^{\pm}}}).
\end{eqnarray}
In the above $\beta_{\pm}$ is the charged Higgs mixing angle. For the expression of the function $G$ and other partial decay widths  of $H^{\pm}$ into two fermion, gauge bosons, see \cite{Aoki:2011pz}. We show the branching ratio of $H^{\pm \pm}$ and $H^{\pm}$ in Fig.~\ref{f:branching}, for two benchmark values of doubly charged Higgs mass,  $M_{H^{\pm \pm}}=247.3$ GeV and $M_{H^{\pm \pm}}=894.02$ GeV, respectively.
\begin{figure}[t]
\begin{center}
\includegraphics[scale=0.45]{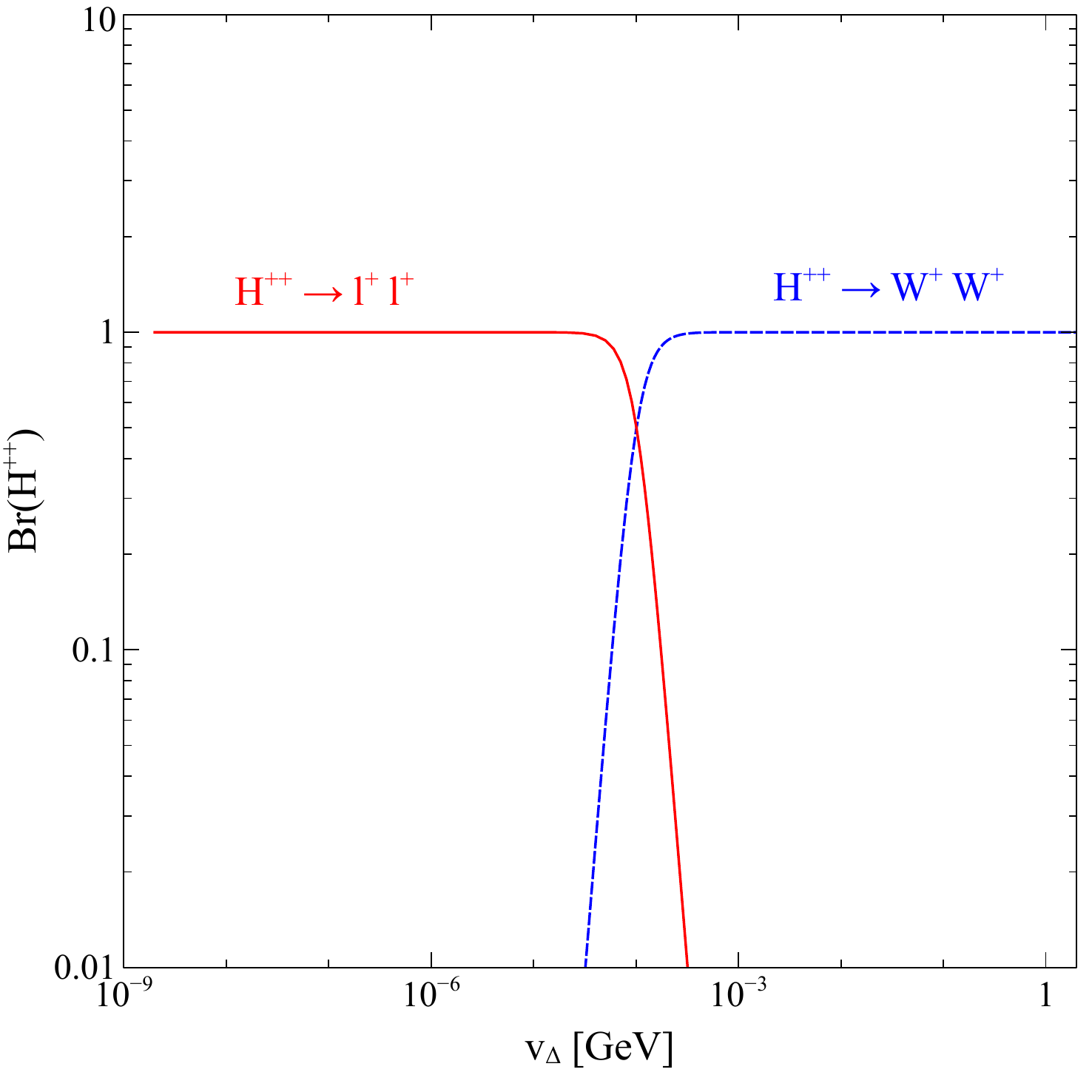}
\includegraphics[scale=0.45]{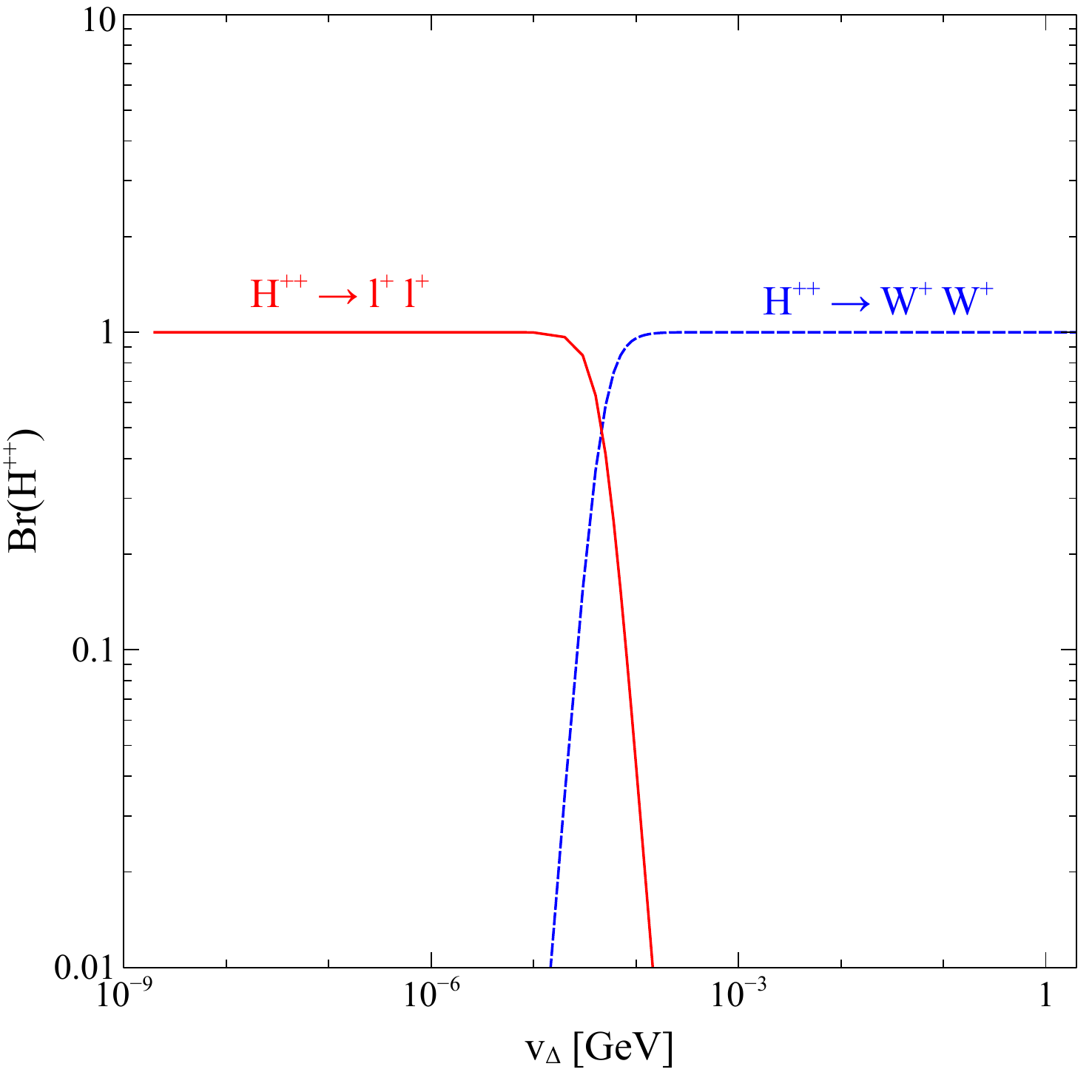}
\includegraphics[scale=0.45]{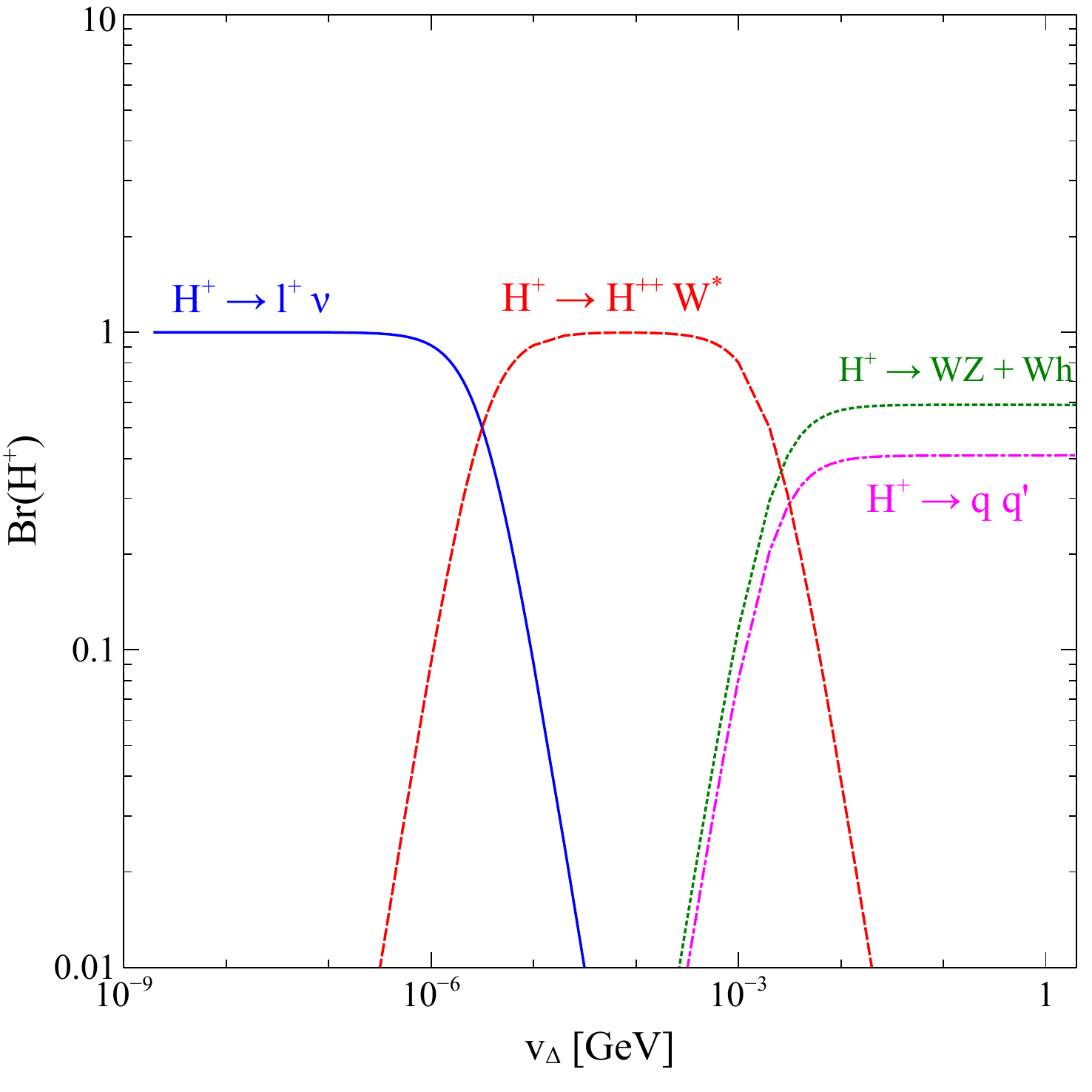}
\includegraphics[scale=0.45]{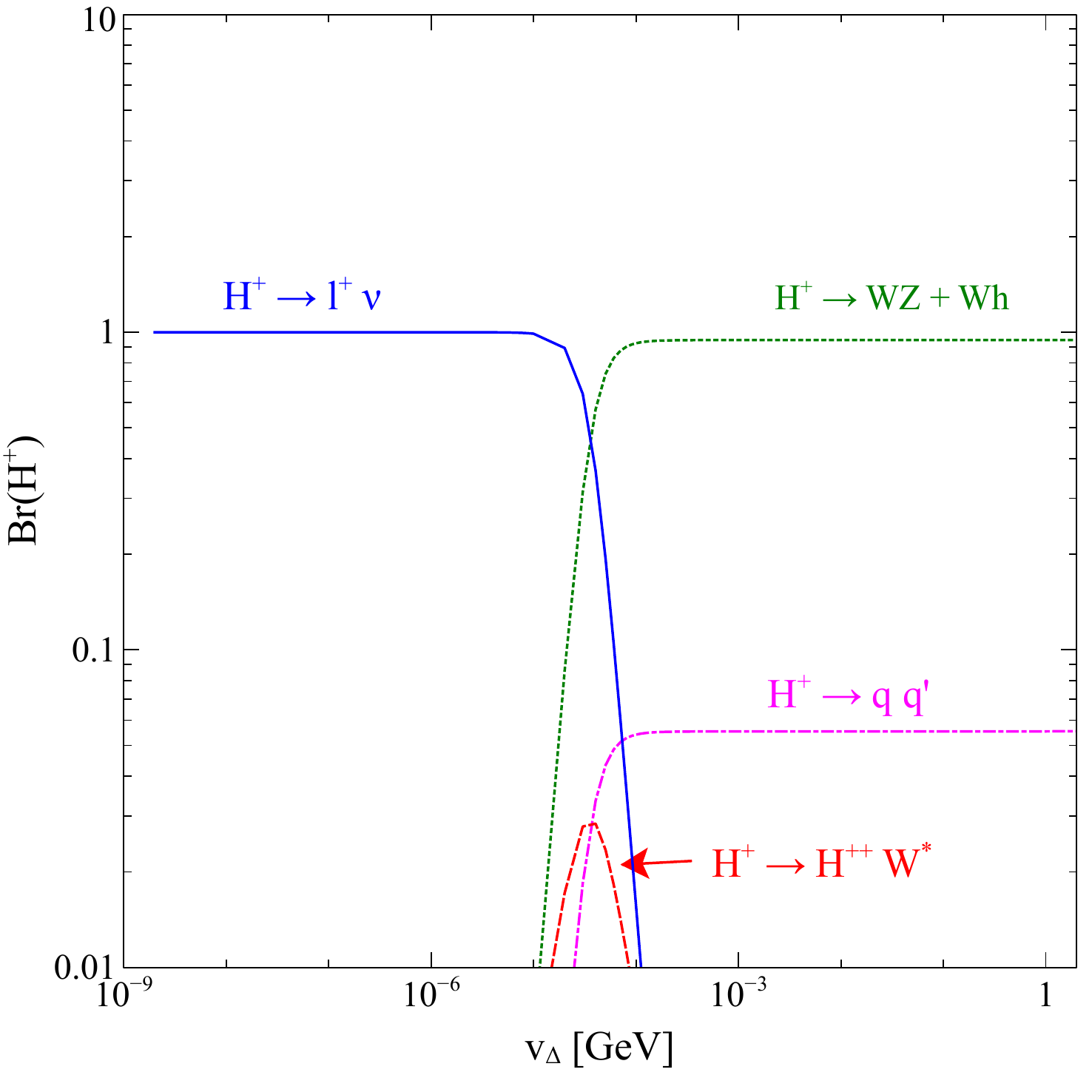}
\caption{Upper panel: the  branching ratios of $H^{\pm \pm}$  for masses $M_{H^{\pm \pm}}= 247.30$ GeV and $M_{H^{\pm \pm}}= 894.02$ GeV.
Lower panel: branching ratios  of $H^{\pm}$ for the  masses, $M_{H^{\pm}}= 250.35$ GeV and $M_{H^{\pm}}= 894.89$ GeV. The other relevant parameters are kept fixed at
$\lambda_i = 0.1$ (for $i$ = 1 to 4) and $\lambda = 0.52$.}
\label{f:branching}
\end{center}
\end{figure}
In the upper panel of Fig.~\ref{f:branching}, we show  the variation of the branching ratios  of doubly charged Higgs
boson for the  two chosen benchmark mass points.  The lower panel shows the variation of the branching ratio  of singly charged Higgs $H^{\pm}$  into different channels. The lower panel has a different response with the
increase of the doubly charged Higgs mass,  which also implies the increase  of the masses of different other charged and neutral Higgs states.
From the top panel, this is evident, that there is hardly any change in the branching ratio of doubly charged Higgs for the variation of its mass, except
a slight shift in the overlapping region of the two branching ratios.  On the other hand in the lower panel, the scenario is completely different 
and one can easily see a huge variation in the branching ratio  of the different decay channel of $H^{\pm}$ 
due to the change in mass of the doubly charged Higgs. This happens because with the increase of the 
doubly charged Higgs mass the ratio $\frac{M_{H^{\pm\pm}}}{M_{H^{\pm}}} \rightarrow 1$, hence the
decrement in the decay width of $H^{+} \rightarrow H^{++} W^{-\,*}$ channel occurs due to the phase space
suppression.

\begin{figure}[t]
\begin{center}
\includegraphics[scale=0.5, angle=90,origin=c]{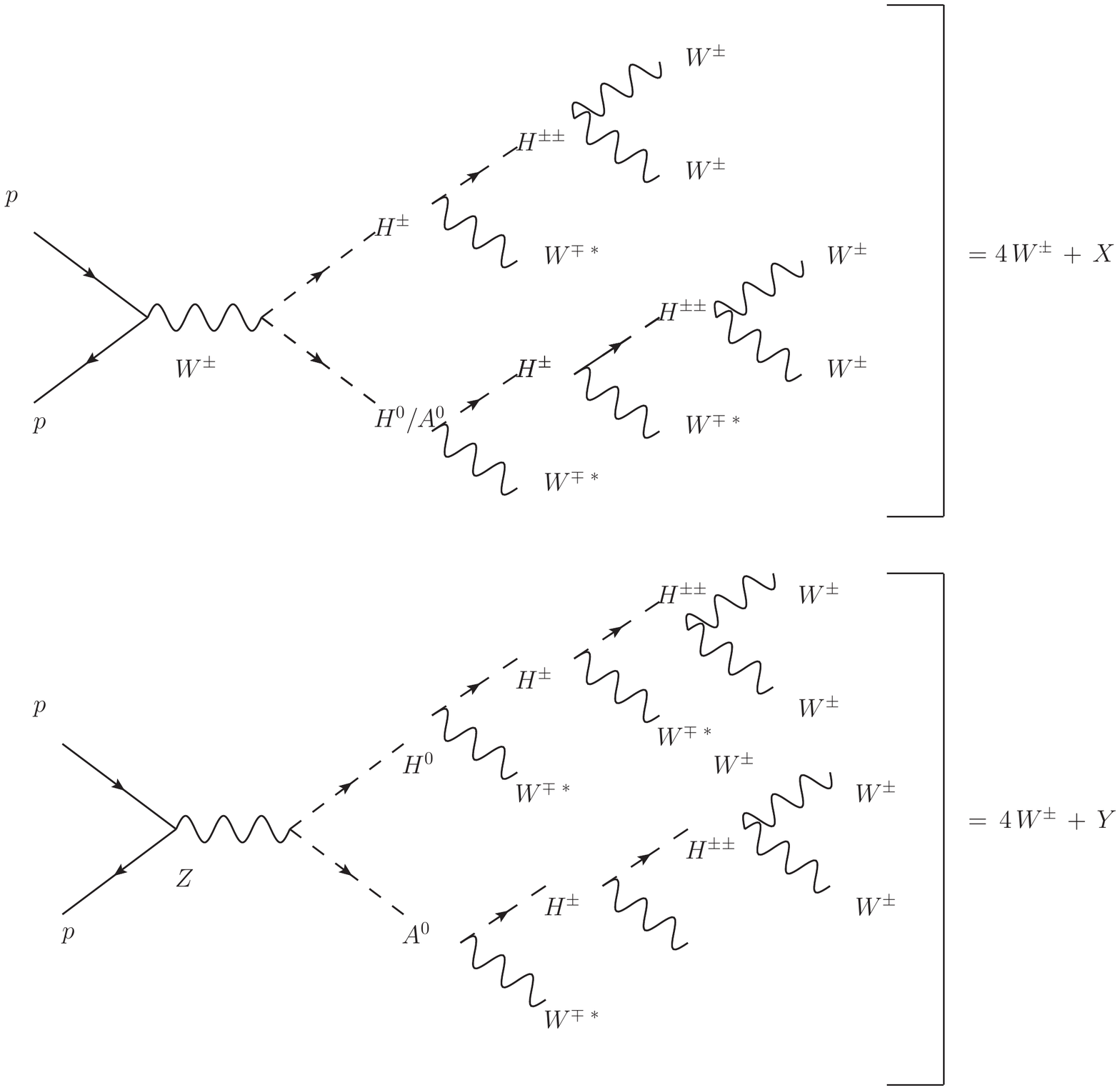}
\caption{Feynman diagrams for $p p \to H^{\pm} H^0/A^0$,  $p p \to H^0 A^0$, and the subsequent decays of  $H^0/A^0 \to H^{\pm} {W^{-}}^*$, $H^{\pm} \to H^{\pm \pm} {W^{-}}^*$, and $H^{\pm \pm} \to W^{\pm} W^{\pm}$. }
\label{f:feynman}
\end{center}
\end{figure}

\begin{figure}[t]
\begin{center}
\includegraphics[scale=0.5, angle=0]{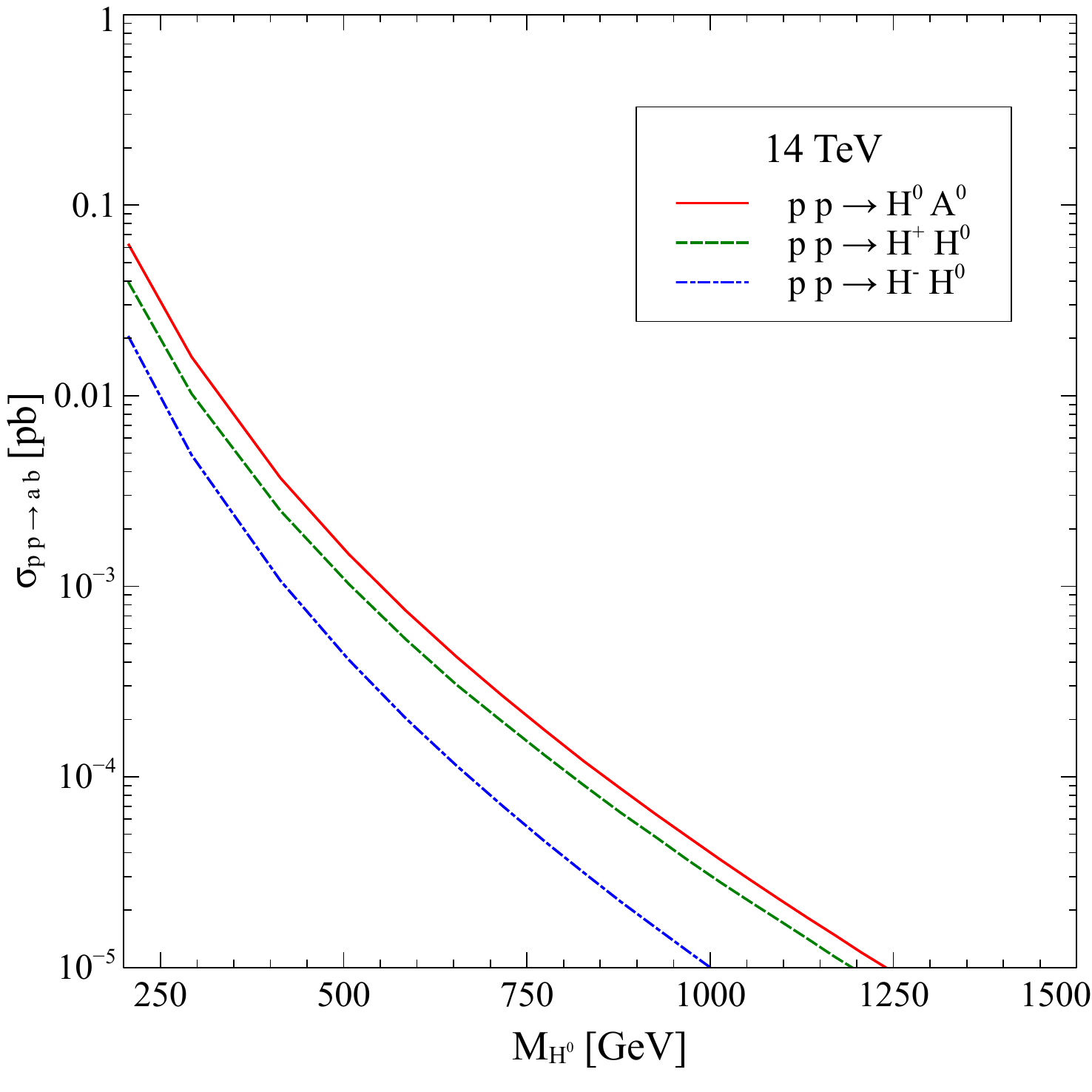}
\includegraphics[scale=0.5]{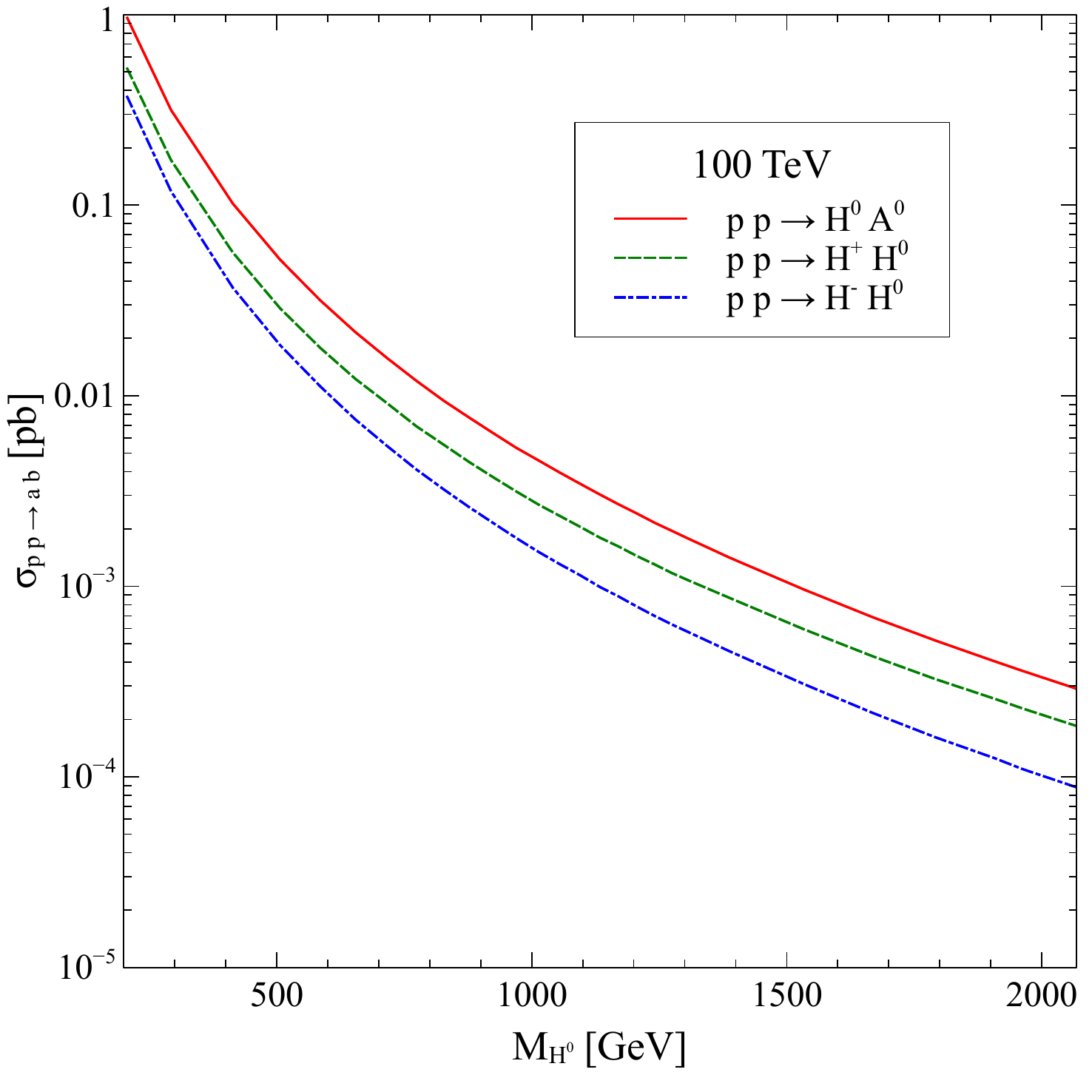}
\caption{Left panel: cross-section for associated production $H^{\pm} H^0$, $H^{0} A^{0}$ vs the mass of $H^{\pm}$.
The c.m.energy is $\sqrt{s}=14$ TeV. The other parameters are fixed at $\lambda_{i} = 0.1$ ($i$ $= 1$ to $4$), $\lambda = 0.52$, $v_{\Delta} = 10^{-3}$ GeV and $\mu$ has been varied between 
$2 \times 10^{-3}$ GeV to $4.5 \times 10^{-2}$ GeV to
vary the mass of the particles. The production cross section for $H^{+}A^{0}$ and $H^{-}A^{0}$
are same with $H^{+}H^{0}$ and $H^{-}H^{0}$, respectively. Right panel: The same plot for  higher c.m.energy $\sqrt{s}=100$ TeV.  }
\label{f:cross}
\end{center}
\end{figure}

A number of searches have been proposed at the LHC to discover   $H^{\pm \pm}$ using multilepton signatures. The  searches in    \cite{ Perez:2008ha, Melfo:2011nx, delAguila:2008cj, Mitra:2016wpr} focussed on 
 the pair and associated production with the $H^{\pm\pm}$ decaying into leptonic, gauge boson states.   Below we discuss the existing constraints on  $H^{\pm \pm}$ from LEP and LHC searches.

\begin{itemize}

\item

{{ Constraint from LEP-II}}: The search for doubly-charged Higgs boson $H^{\pm \pm}$ decaying into charged leptons have been performed at LEP-II. This constrains the mass  $M_{H^{\pm \pm}} > 97.3$ GeV  \cite{Abdallah:2002qj} at 95$\%$ C.L.

\item

{Constraints from pair and associated production}:  Stringent constraint on  $M_{H^{\pm \pm}}$  have been placed by  the 13 TeV LHC searches. These searches analysed  $H^{\pm \pm } \to l^{\pm } l^{\pm}$ channel. The CMS collaboration looked for different leptonic flavors including $e e, e \mu, e \tau, \mu \mu, \mu \tau $ 
and $\tau \tau$. In addition, the CMS searches also include the associated production $p p  \to H^{\pm \pm} H^{\mp}$ and the subsequent decays, $H^{\pm} \to l^{\pm} \nu$. This combined channel of pair-production and associated production gives the stringent constraint $M_{H^{\pm \pm}} > 820$ GeV \cite{CMS-PAS-HIG-16-036} 
at $95 \%$ C.L for $e, \mu$ flavor. The realistic bound depends on the neutrino mass matrix~\cite{Chun:2003ej}.  Similar  constraint from ATLAS searches have been placed on the mass of doubly charged Higgs, that takes into account only  pair-production. The bound is $M_{H^{\pm \pm}} > 870 $ GeV at $95 \%$ C.L \cite{Aaboud:2017qph}. 
Note that these limits are valid only for a small triplet vev $v_{\Delta} < 10^{-4}$ GeV.  {Additionally,   ATLAS  looked into the pair-production of doubly charged Higgs, with  subsequent decays 
into gauge bosons, resulting in multi-lepton final states.  The search in 
\cite{Aaboud:2018qcu}, have constrained the mass of doubly charged Higgs $M_{H^{\pm \pm}}$ in between 200-220 GeV at 95$\%$ C.L. This is valid for the triplet vev $v_{\Delta} > 10^{-4}$ GeV, where the 
gauge boson decay is most dominant. }

\item

Constraint from VBF: For larger values of the triplet vev {$v_{\Delta} > 10^{-4}$ GeV}, the leptonic branching ratio {becomes smaller}. Instead the {decay} mode  $H^{\pm \pm} \to W^{\pm } W^{\pm}$ is dominant. Therefore the searches in vector boson fusion (VBF) become more important. A search for $p p  \to j j H^{\pm \pm} \to j j  W^{\pm } W^{\pm}$ at the 8 TeV LHC in the VBF channel sets a constraint on the triplet vev $v_{\Delta} \sim 25$ GeV for $M_{H^{\pm \pm} } \sim 300$ GeV \cite{Khachatryan:2014sta}. This constraint has been updated  \cite{Sirunyan:2017ret} using 13 TeV data at the LHC. Such a large triplet vev is anyway excluded by the $\rho$ parameter bound~\cite{Chun:2012jw} in the minimal type-II seesaw model.

\end{itemize}

The above mentioned constraints imply  that a large range of triplet vev $v_{\Delta} > 10^{-4}$ GeV exists, where  low mass  of $M_{H^{\pm \pm}} > 220$ GeV is still allowed. For lower triplet vev $v_{\Delta} < 10^{-4}$ GeV, the mass constraint is more conservative $M_{H^{\pm \pm}} > 870$ GeV.  In our analysis of tetra-lepton signatures, we therefore choose both the
lighter and heavier mass points. 

\section{Large triplet vev and same-sign tetra-lepton signature for $\sqrt{s}=14$ TeV}
\label{same sign tetralepton section}

We explore the tetra-lepton signature arising from a lighter charged Higgs and neutral Higgs decay. 
We consider  associated production  of  $H^{\pm}$ along-with $H^0/A^0$. For triplet vev in between $10^{-5}\, \rm{GeV} < v_{\Delta} < 10^{-3}\, \rm{GeV}$, and assuming  mass hierarchy between singly and doubly charged Higgs  $M_{H^{\pm}}> M_{H^{\pm \pm}}$, the cascade decay of $H^{\pm}$ into $H^{\pm \pm} W^*$ is predominant.  In the same triplet vev region, $H^{0}/A^{0} \to H^{\pm} W^*$ decay is  also significantly large. We furthermore consider the gauge boson decay modes of $H^{\pm \pm} \to W^{\pm} W^{\pm}$, that has  large 
branching ratio  for  $v_{\Delta} > 10^{-4}$ GeV and subsequent leptonic decay of  the produced on-shell $W^{\pm}$.  For the signal, therefore,  the complete process is~\cite{Chun:2012zu},

\begin{itemize}
\item
 $p p  \to H^{\pm} H^{0}/ H^{\pm} A^0 \to H^{\pm \pm} {W^{\pm *}}  H^{\pm} {W^{\mp *}} \to  H^{\pm \pm} {W^{\pm *}}  H^{\pm \pm} {W^{\mp *}} {W^{\mp *}} \to 4W^{\pm}+ X$
 \item
 $p p  \to H^0 A^0 \to  H^{\pm} {W^{\mp *}} H^{\pm} {W^{\mp *}} \to  H^{\pm \pm} {W^{\pm *}}  H^{\pm \pm} {W^{\mp *}} {W^{\mp *}}  {W^{\mp *}} \to 4W^{\pm}+ Y$
 \end{itemize}

The Feynman diagrams for these above two processes have been shown in Fig.~\ref{f:feynman}.  Note that this phenomenon of wrong sign leptons production occurs as $\Delta^0$ can oscillate to $\Delta^{0\dagger}$ and vice versa. As a result, $H^0$ and $A^0$, sharing the same final states, can mix together like in the $B^0-\overline{B^0}$ system. Finally we can write the cross-section for these signals as:

\begin{itemize}
\item
$\sigma(p p \to H^{\pm} H^0/A^0) \times F_1 \times \text{Br}( H^{\pm} \to H^{\pm\pm} {W^-}^*)^2\times \text{Br}( H^{0}/A^0 \to H^{\pm} {W^-}^*)\times \text{Br}( H^{\pm\pm} \to W^{\pm} W^{\pm})^2$

\item

$\sigma(p p \to H^{0} A^0) \times F_2 \times \rm{Br( H^{0}/A^0 \to H^{ \pm} {W^-}^*)^2} \times \rm{Br( H^{\pm} \to H^{\pm \pm} {W^-}^*)}^2 \times \rm{Br( H^{\pm \pm} \to W^{\pm} W^{\pm})}^2$

\end{itemize}

 In the above  $F_{1,2}$ are 
\begin{eqnarray}
F_1=\frac{x^2}{1+x^2},  ~~~  F_2=\frac{2+x^2}{2(1+x^2)} \times \frac{x^2}{2(1+x^2)}, ~~ \textrm{with}\, x=\frac{\delta M}{\Gamma_{H^0/A^0}}=\frac{M_{H^0}-M_{A^0}}{\Gamma_{H^0/A^0}}
\end{eqnarray}
 When the two decay widths $\Gamma_{H^0}$ and $\Gamma_{A^0}$ are nearly equal, i.e., $\Gamma_{H^0} \simeq \Gamma_{A^0}$. The generatisation of these two processes to the case of $\Gamma_{A^0} \neq \Gamma_{H^0}$ is 

\begin{itemize}
\item
$\sigma(p p \to H^{\pm} H^0/A^0) \times G_1 \times \rm{Br( H^{\pm} \to H^{\pm \pm} {W^-}^*)}^2\times \rm{Br( H^{0}/A^0 \to H^{ \pm} {W^-}^*)}\times \rm{Br( H^{\pm \pm} \to W^{\pm} W^{\pm})}^2$

\item

$\sigma(p p \to H^{0} A^0) \times G_2 \times \rm{Br( H^{0}/A^0 \to H^{ \pm} {W^-}^*)^2} \times \rm{Br( H^{\pm} \to H^{\pm \pm} {W^-}^*)}^2 \times \rm{Br( H^{\pm \pm} \to W^{\pm} W^{\pm})}^2$

\end{itemize}

where $G_1$ and $G_2$ have the following forms:

\begin{eqnarray} 
G_1=\frac{x^2+y^2}{2(1+x^2)}, G_2=\frac{2+x^2-y^2}{2(1+x^2)} \times \frac{x^2+y^2}{2 (1+x^2)}, ~~~ {\rm{with}}  ~~~~ x=\frac{\delta M}{\Gamma},\nonumber\\
\Gamma=\frac{\Gamma_H^0+\Gamma_A^0}{2} , ~~ \textrm{and} ~~ y=\frac{\Gamma_H^0-\Gamma_A^0}{\Gamma_H^0+\Gamma_A^0}
\end{eqnarray}

Note that, to compute the tetra-lepton signature, one needs to take into account the leptonic branching ratios from $W$. In our analysis, we consider both the $W \to l \nu$, with $l=e, \mu$, as well as $W \to \tau \nu$, with the 
leptonic decays of $\tau$ included.
To compute the cross-section, we  implement the model in 
FeynRules(v2.3) \cite{Alloul:2013bka}. The  UFO output   is then fed  into  MadGraph5\_aMC@NLO(v2.6) \cite{Alwall:2014hca}   that  generates  the parton-level  events. We use the default pdf 
NNPDF23LO1 \cite{Ball:2013hta} for computation.  We perform parton showering and hadronization  with Pythia8 \cite{Sjdostrand:2007gs} and  analyse the HepMC \cite{Dobbs:2001ck} event files.  The above cross-sections $p p \to H^{\pm} H^0$ and 
$p p  \to H^0 A^0$ depend on the masses of the neutral and charged Higgs. We therefore show the variation of associated production cross-section  of $p p \to H^{\pm} H^0/A^0$ and $p p \to H^0 A^0$  with the mass of $H^0$  in Fig.~\ref{f:cross}.  For c.m.energy $\sqrt{s}=14$ TeV, the cross-section for $p p  \to H^0 A^0$ varies in between  $1-70 $ fb, for neutral Higgs mass between $200-500$ GeV.  For $p p \to H^{+} H^0/A^0$, the cross-section is very similar, only lower than by a factor of $\mathcal{O}(1.5)$.  For $p p \to H^{-} H^0/A^0$, cross-section is smaller due to the parton distribution function.  In addition, we also show the production cross-section for 
a future $p p $ collider, with c.m.energy $\sqrt{s}=100$ TeV.  As is evident from the right panel of Fig.~\ref{f:cross},  the production cross-section is quite large for higher c.m.energy, and   
multi-TeV  Higgs mass  can be probed.

Note that the production cross-sections for $p p \to H^{\pm} H^0/A^0$ depends  on both the parameters $\lambda_4$ and the triplet vev $v_{\Delta}$. For a fixed value of 
$\mu$, the triplet vev  primarily 
governs the masses of the Higgs $H^{\pm}, H^{0}/A{^0}$, while the parameter $\lambda_4$ determines their mass difference. In the left panel of Fig.~\ref{f:cross HpH0 and br}, we show the variation of production cross section for the process $p p \rightarrow H^{+} H^{0}$ in the
$v_{\Delta}-\lambda_{4}$ plane for a benchmark value  of  neutral Higgs,
$M_{H^{0}}  \sim 253$ GeV. {For the process $p p \rightarrow H^{-} H^{0}$,  the plot is very similar, only the production cross section is relatively smaller by a factor of two.  The channel $p p \to H^0 A^0$ has the largest cross-section, 
larger than $p p \to H^{+} H^0$ by almost  a factor of $\mathcal{O}(1.4-1.7)$. Since $\lambda_4$  has a very nominal effect  on the mass splitting of $H^0, A^0$, the cross-section of this channel is almost fixed 
in the entire plane of $\lambda_4 - v_{\Delta}$, and thus does not vary.   }



\begin{figure}[t]
\begin{center}
\includegraphics[scale=0.18]{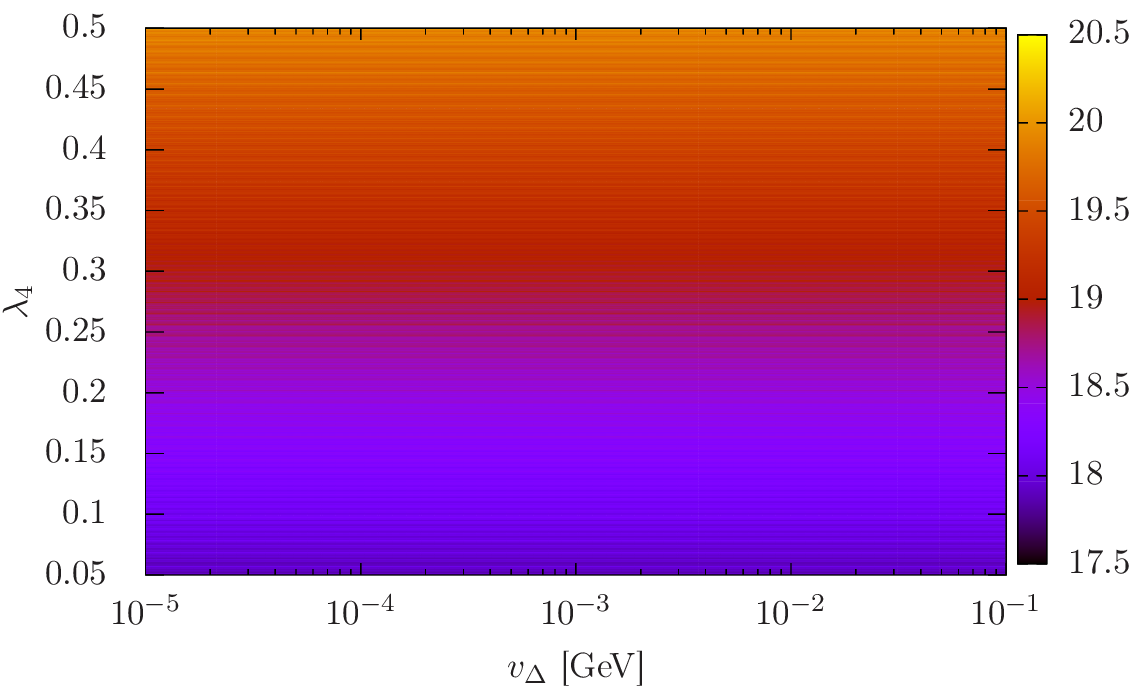}
\includegraphics[scale=0.18]{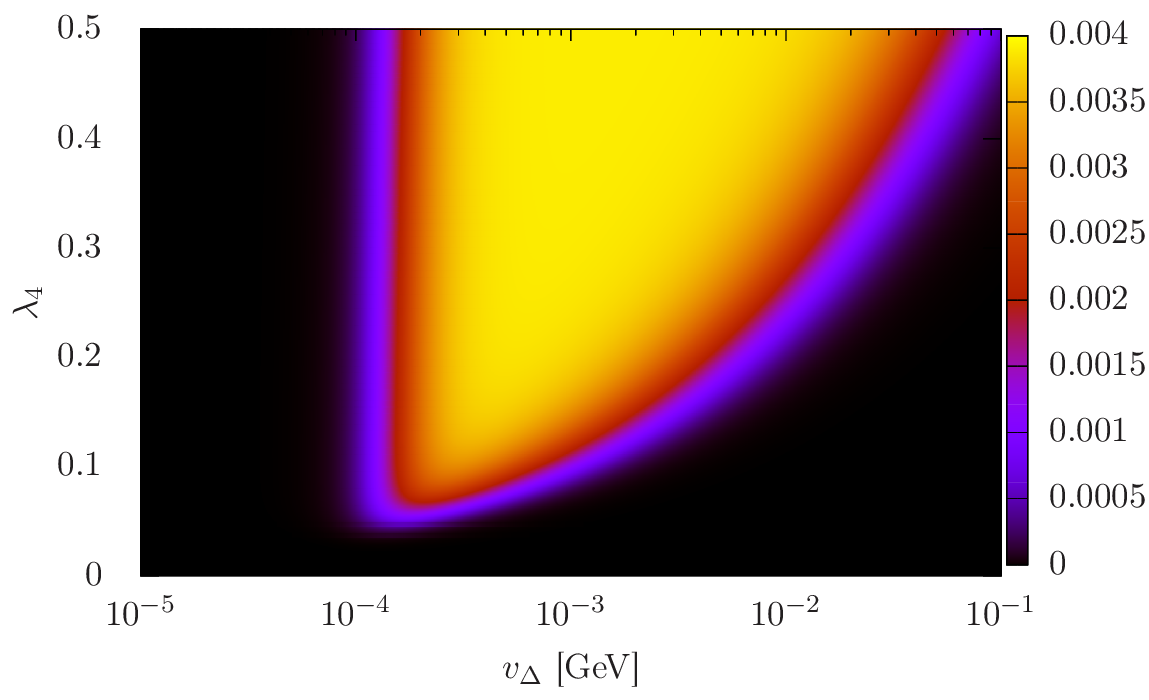}
\caption{ Left panel: cross section in fb for the channel $p p  \to H^{+}H^0$~ for  the c.m.energy $\sqrt{s}=14$ TeV~($M_{A^0}=253$ GeV). Right panel: 
Product of branching ratios $\text{Br}( H^{\pm} \to H^{\pm\pm} {W^-}^*)^2\times \text{Br}( H^{0}/A^0 \to H^{\pm} {W^-}^*)\times \text{Br}( H^{\pm\pm} \to W^{\pm} W^{\pm})^2 \times\text{Br}(W^{\pm}\to\ell\nu)^4$ for the process $pp\to H^{\pm}H^0/A^0$ with mass of $A^0$ being fixed as $M_{A^{0}}=253$~GeV. For the second process $pp\to H^0/A^0$, behaviour of the product of branching ratio will be same.} 
\label{f:cross HpH0 and br}
\end{center}
\end{figure}

{The doubly, singly charged, and neutral Higgs bosons  will decay through a number of subsequent decay modes, leading to the same-sign tetra-lepton final states.  The two key parameters are again triplet vev 
$v_{\Delta}$ and 
the coupling $\lambda_4$. 
Since a number of branching ratios are involved in the same-sign tetra-lepton process, we show the product of these branching ratios.    }
In the right panel of Fig.~\ref{f:cross HpH0 and br}, we  show  the variation of the product of branching ratios $\text{Br}( H^{\pm} \to H^{\pm\pm} {W^-}^*)^2\times \text{Br}( H^{0}/A^0 \to H^{\pm} {W^-}^*)\times \text{Br}( H^{\pm\pm} \to W^{\pm} W^{\pm})^2 \times\text{Br}(W^{\pm}\to\ell\nu)^4$ for the process $pp\to H^{\pm}H^0/A^0$ in the $v_{\Delta} - \lambda_{4}$ plane. From  the top panel of Fig.~\ref{f:branching},  it is evident 
that the doubly charged Higgs $H^{\pm\pm}$ decays predominantly to same
sign $W^{\pm}W^{\pm}$ state.  For smaller range of the triplet vev
it entirely decays to $l^{\pm}l^{\pm}$ final state. This is reflected in 
Fig.~\ref{f:cross HpH0 and br}, where  there is a sharp change in branching ratio  around  $10^{-4}$ GeV. 
The product goes to zero in the left side of this line (as shown by the black region). 
In the right side of this line, the product can be large,   as indicated by the colour bar. We stress that, the product of the branching  ratios has a significantly large  value for a wide range of the triplet vev, 
$10^{-4}\,{\rm GeV} < v_{\Delta} <10^{-2}\,{\rm GeV}$. Therefore, in this region, there will be handful of events for same-sign tetra-lepton final states, that can be tested at LHC.  In the  next section, we will see how  this large range of triplet vev  shrinks
to a very narrow  range for higher masses of the charged  and neutral Higgs. This occurs due to 
significant change in  branching ratios of the channel $H^{\pm} \to H^{\pm\pm} {W^-}^*$ for the same value of $\lambda_4$.

\begin{figure}[t]
\begin{center}
\includegraphics[scale=0.18]{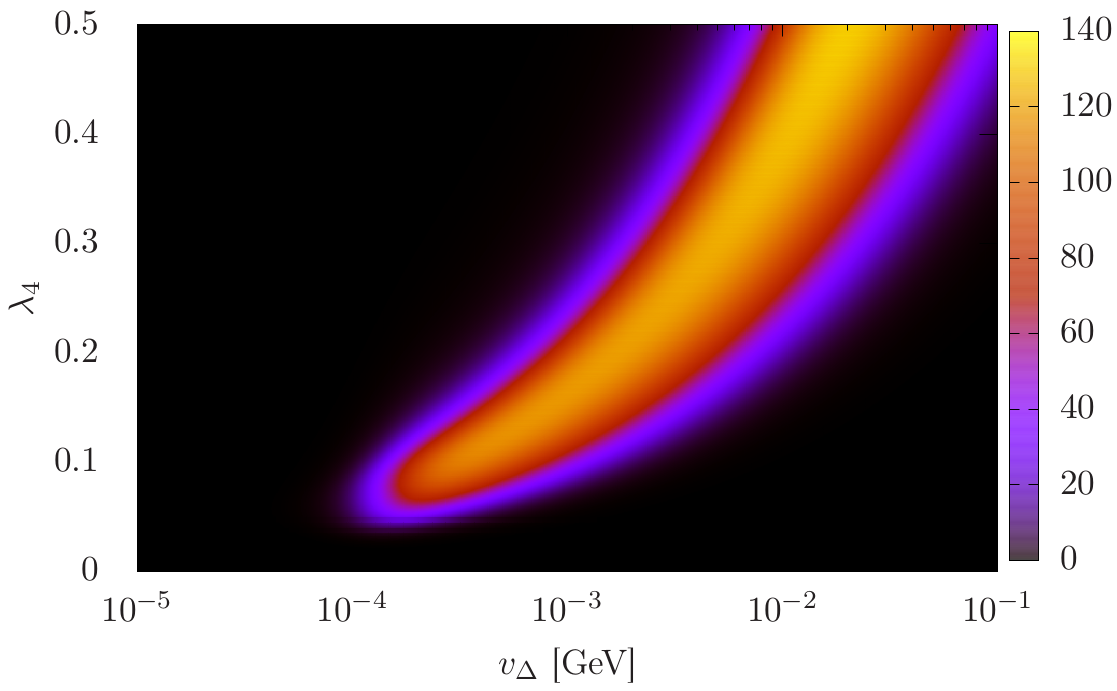}
\includegraphics[scale=0.18]{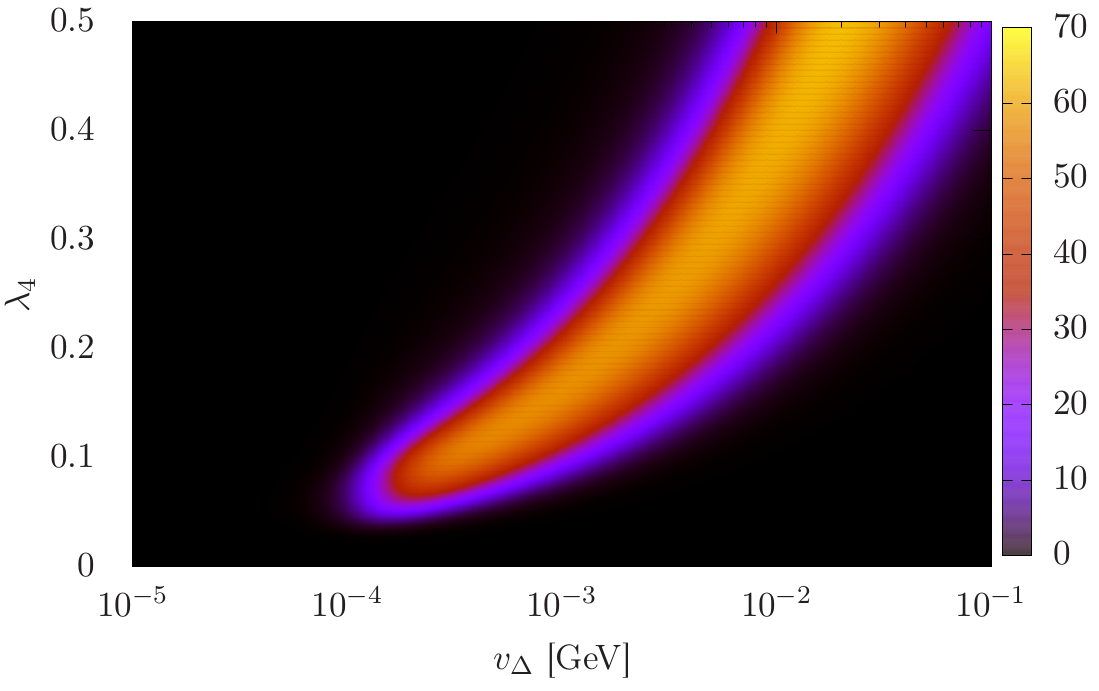}
\includegraphics[scale=0.18]{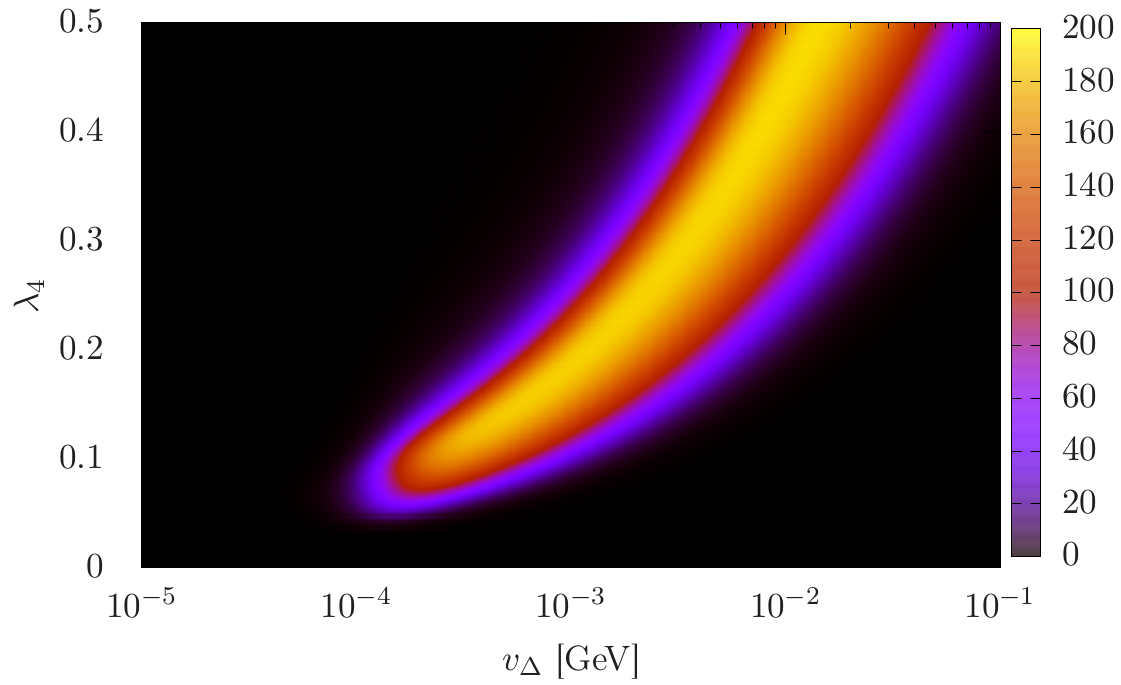}
\caption{This plot represents   number of same-sign tetralepton events for  mass $M_{H^0} \sim M_{A^0}=253$~GeV. Left figure of upper panel: number of same-sign tetralepton events $\ell^+\ell^+\ell^+\ell^+ + X$ from $p p \to H^{+}H^0/A^0$ followed by  subsequent decays of $H^{+}, H^0, A^0$. Right figure of upper panel: number of same-sign tetralepton events $\ell^-\ell^-\ell^-\ell^- + X$ from $p p \to H^{-}H^0/A^0$ and subsequent decays. Lower panel: number of same-sign tetralepton events $\ell^+\ell^+\ell^+\ell^+ + Y$ or $\ell^-\ell^-\ell^-\ell^- + Y$ from $p p \to H^{0}A^0$ and subsequent decays. For the doubly charged Higgs, we consider $H^{\pm \pm} \to W^{\pm} W^{\pm}$ decay mode.  The c.m.energy $\sqrt{s}=14$ TeV and we consider luminosity $\mathcal{L}=3000$ $\rm{fb}^{-1}$. For this range of $\lambda_4$, the masses of $H^{\pm}$ and $H^{\pm \pm}$ varies at most by $M_{H^0}-M_{H^{++}}\sim 32$ GeV and $M_{H^0}-M_{H^{+}}\sim 15$ GeV respectively.}
\label{event number for mass 253}
\end{center}
\end{figure}

In Fig.~\ref{event number for mass 253}, we show  the variation of number of events for the same-sign tetra-lepton signature, where 
we consider integrated luminosity $\mathcal{L}=3000 $ $\rm{fb}^{-1}$. This has been obtained by folding the production cross-section with the overall branching ratio, and integrated luminosity. We also implement 
few basic cuts at the {pythia} 
level. These are {$ p_{T}({e^{\pm}/\mu^{\pm}} ) > 10 \, \textrm{GeV}$, $ |\eta ({e^{\pm}/\mu^{\pm}})| < 2.5 $. We obtain  a cut-efficiency $c_{eff}=0.62$ for  $ M_{H^{0}} = 253\,$GeV}, that we include in our  calculation of  total number of events.  We consider the processes $p p \to H^{+}H^0/A^0$ (top left),
$p p \to H^{-}H^0/A^0$ (top right) and $p p \to H^{0}A^0$ (bottom).To calculate  the
number of events we followed the  prescription given  at the beginning of
Section \ref{same sign tetralepton section}.   
As we can see from the bottom left plot of Fig.~\ref{f:branching},  that for the low mass range of the particle spectrum, 
the channel $H^{\pm} \rightarrow H^{\pm\pm} W^{-\,*} $ has $100\%$ branching ratio for a wide range of triplet vev.
Hence in all these three plots, we get a reasonable number of events for  triplet vev  
$v_{\Delta} \sim 10^{-4} - 10^{-1}$ GeV. As exhibited in Fig.~\ref{f:cross}, the cross section
for the different final states have the following hierarchies 
$\sigma(p p \to H^{0}A^0) > p p \to \sigma(H^{+}H^0/A^0)  > \sigma(p p \to H^{-}H^0/A^0)$.  The same hierarchy also translates in  the number of events.
All the three plots have a similar kind of morphology in the $v_{\Delta} - \lambda_{4}$
plane and the nature of the variation of the number of events can be understood in the following way.
Since we are considering $H^{\pm\pm} \rightarrow W^{\pm}W^{\pm}$ channel which start contributing when triplet vev
is $v_{\Delta} > 10^{-4}$ GeV, so the number of events $ N_{evt}> 5$ starts around this region of triplet vev. 
As shown  in Fig.~\ref{f:cross HpH0 and br}, the cross section increases with  larger  $\lambda_{4}$, while 
 the branching ratio for the channel $H^{\pm} \rightarrow H^{\pm\pm} W^{-\,*} $
decreases (bottom right plot of Fig.~\ref{f:branching}) for larger  triplet vev, leading to the specific  variation of the number of events shown in  \ref{event number for mass 253}.

\section{ Inclusive same-sign tetra-lepton signature for $\sqrt{s}=100$ TeV \label{signal at 100 TeV}}

\begin{figure}[t]
\begin{center}
\includegraphics[scale=0.18]{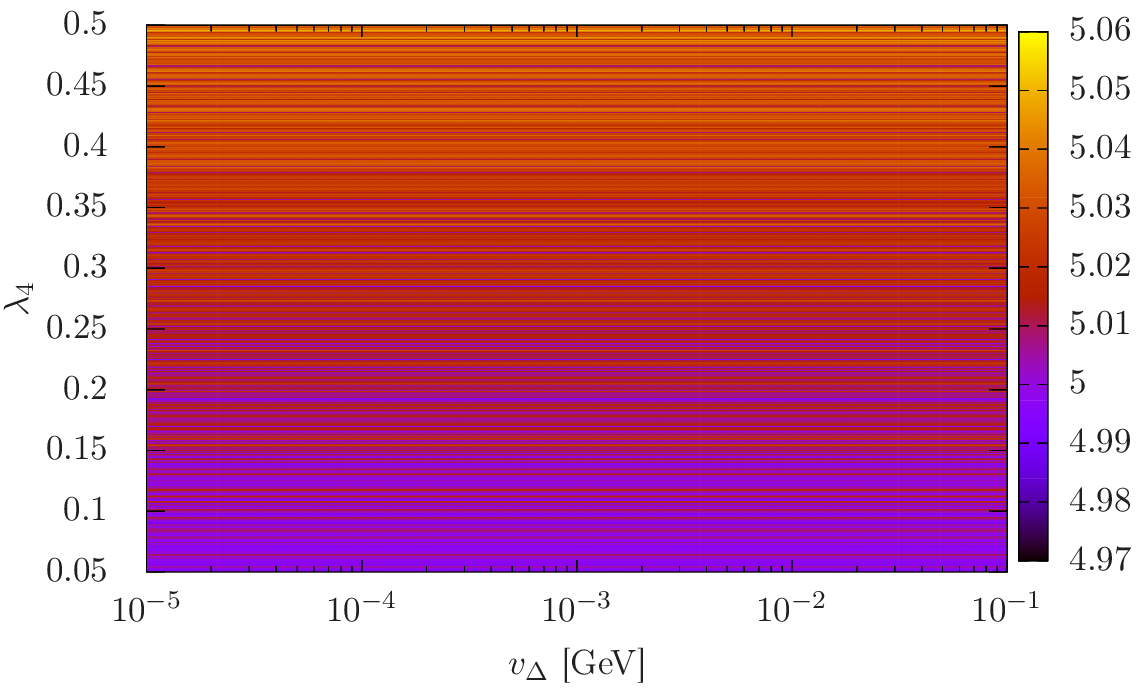}
\caption{ Cross section in fb for the channel $p p  \to H^{+}H^0$ for  the  mass  $M_{A^0}=900$~GeV. We consider c.m.energy  $\sqrt{s}=$100 TeV.} 
\label{f:cross HpH0 850}
\end{center}
\end{figure}

\begin{figure}[t]
\begin{center}
\includegraphics[scale=0.18]{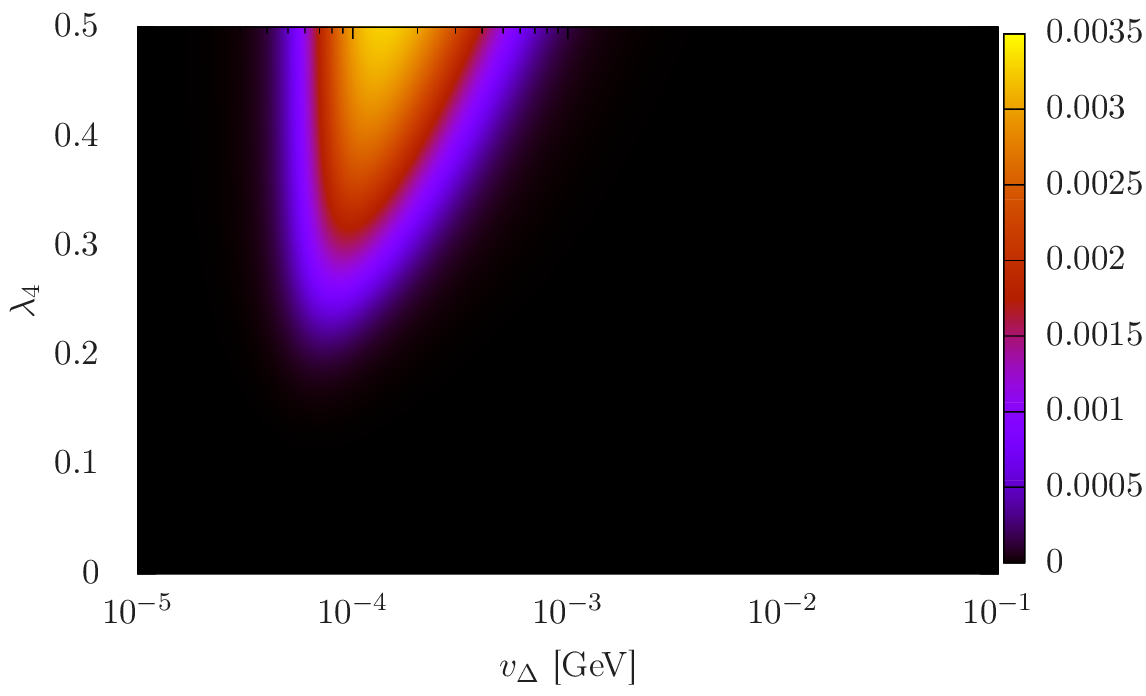}
\includegraphics[scale=0.18]{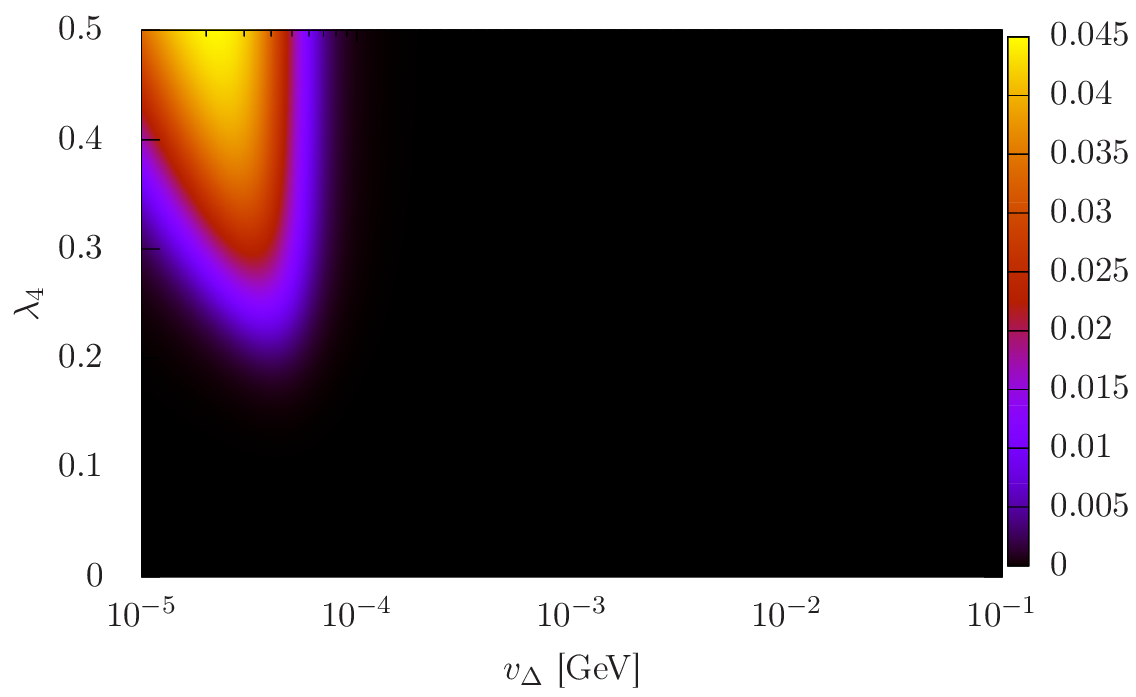}
\includegraphics[scale=0.18]{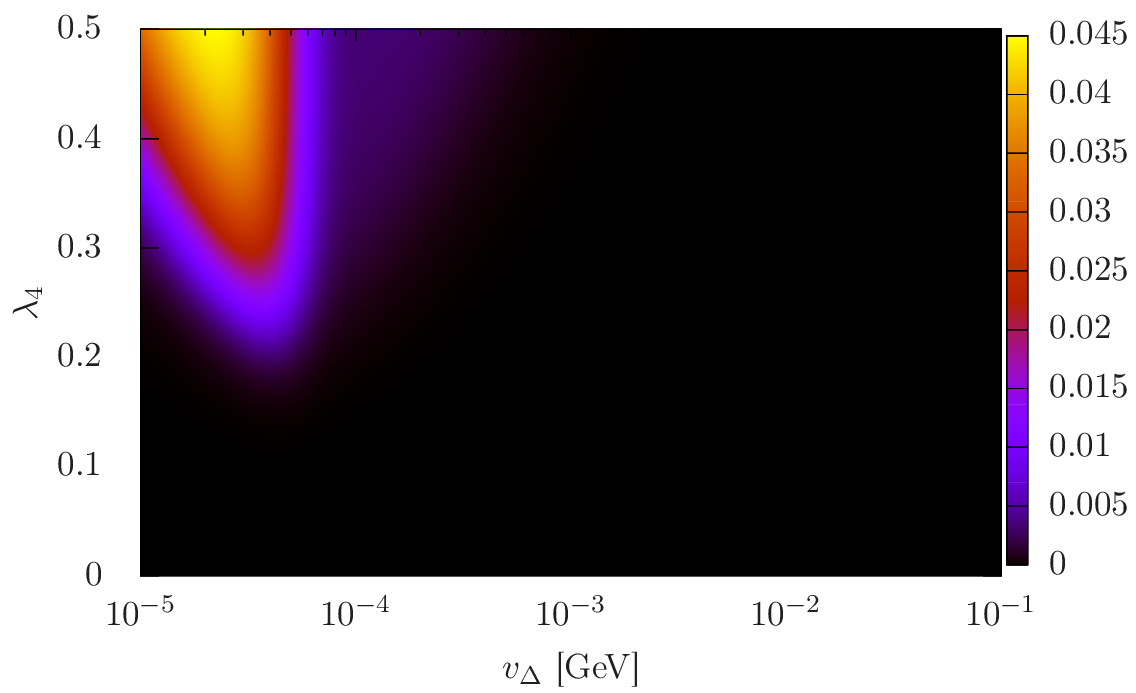}
\caption{Upper panel: this represents the product of branching ratios $\text{Br}( H^{\pm} \to H^{\pm\pm} {W^-}^*)^2\times \text{Br}( H^{0}/A^0 \to H^{\pm} {W^-}^*)\times \text{Br}( H^{\pm\pm} \to W^{\pm} W^{\pm})^2 \times\text{Br}(W^{\pm}\to\ell\nu)^4$~(left figure), $\text{Br}( H^{\pm} \to H^{\pm\pm} {W^-}^*)^2\times \text{Br}( H^{0}/A^0 \to H^{\pm} {W^-}^*)\times \text{Br}( H^{\pm\pm} \to \ell^{\pm} \ell^{\pm})^2$~(right figure). Lower panel:  the sum of these two products of branching ratios for the process $pp\to H^{\pm}H^0/A^0$ with mass of $A^0$ being fixed as $M_{A^{0}}=900$~GeV. For the  process $pp\to H^0 A^0$,  the product of branching ratio is  very similar. }
\label{f:product of BR for HpH0 with 850}
\end{center}
\end{figure}

We consider heavier Higgs, and analyse its discovery prospect at a future $pp$ collider that can operate with c.m.energy $\sqrt{s}=100$ TeV.  Due to the suppression from a number of  branching ratios, observation  of  same-sign tetra-lepton final states will be beyond the scope of 13 TeV LHC. However, this can easily be observed in a future collider with higher c.m.energy. As a benchmark sample, we consider neutral Higgs mass $M_{H^0/A^0}=900$ GeV, and variation of  doubly charged Higgs  of mass  at most 
by $ 5 $ GeV from $M_{H^0/A^0}$.  The chosen value of the doubly charged Higgs mass is consistent with the constraints from 13 TeV LHC searches for the entire range of triplet vev $v_{\Delta} \sim 10^{-9} -1 $ GeV.  Near the triplet
vev $v_{\Delta} \sim 10^{-4}$, both the di-lepton and gauge boson modes will substantially contribute. We therefore  cover a large range of triplet vev $v_{\Delta}$, and consider  the doubly charged Higgs decaying into both the same-sign di-lepton, and gauge boson modes. Hence, in addition to the gauge bosons,   discussed in Sec.~\ref{same sign tetralepton section}, the total cross-section also contains the following contribution from di-lepton decay mode,

\begin{itemize}
\item
$\sigma(p p \to H^{\pm} H^0/A^0) \times G_1 \times \rm{Br}( H^{\pm} \to H^{\pm \pm} W^{-*})^2\times \rm{Br}( H^{0}/A^0 \to H^{\pm} W^{-*})\times \rm{Br}( H^{\pm \pm} \to \ell^{\pm} \ell^{\pm})^2$

\item

$\sigma(p p \to H^{0} A^0) \times G_2 \times \rm{Br}( H^{0}/A^0 \to H^{ \pm} W^{-*})^2 \times \rm{Br}( H^{\pm} \to H^{\pm \pm} W^{-*})^2 \times \rm{Br}( H^{\pm \pm} \to \ell^{\pm} \ell^{\pm})^2$

\end{itemize}

In the above, $l=e, \mu, \tau$, and we finally consider the leptonic branching ratios of $\tau$, while calculating the number of events. 
The functions $G_{1,2}$ have been described in Section.~\ref{same sign tetralepton section}. We show the variation of cross-section in Fig.~\ref{f:cross HpH0 850}.  The cross-section for the mass $M_{A^0}=900$~GeV varies around 5 fb. We next show the variation of  the product of branching ratios in  Fig.~\ref{f:product of BR for HpH0 with 850}  
for  heavier charged and neutral Higgs. For triplet vev smaller than $v_{\Delta} < 10^{-4}$ GeV, the doubly charged Higgs $H^{\pm \pm} \to l^{\pm} l^{\pm}$ is dominant, while around $10^{-4}$ GeV, both the gauge boson
mode and di-lepton are dominant.  For a heavier  singly charged Higgs, the branching ratio  for  $H^{\pm} \to H^{\pm\pm} {W^-}^*$
decay channel is large for a large value of $\lambda_4$.  Note that, for $\lambda_4 \sim 0.1$, the branching ratio becomes  more than $1\%$  in a very small range of the triplet vev (see  Fig.~\ref{f:branching}).  This in turns has large  effect on the total branching ratio, and 
is clearly visible in all the three plots
of Fig.~\ref{f:product of BR for HpH0 with 850}.  The region in $v_{\Delta}$, in which the overall branching ratio is larger than $0.5\%$ is   now considerably smaller.  
The left plot in the top  panel  represents the overall branching ratio with only $H^{\pm \pm} \to W^{\pm} W^{\pm}$ decay included. The plot in the right panel shows the total branching ratio for 
$H^{\pm \pm} \to l^{\pm} l^{\pm}$.  The product of the branching ratio is smaller  for the case of 
$H^{\pm\pm} \rightarrow W^{\pm} W^{\pm}$ due to additional suppression from $\text{Br}(W^{\pm} \rightarrow \ell^{\pm} \nu)^4$. In the lower panel, we show  the sum of these two branching ratios. 
 The higher values of the product of the branching ratios is governed by  $H^{\pm\pm} \rightarrow l^{\pm} l^{\pm}$  
decay mode (relevant for $v_{\Delta} \lsim 10^{-4}$ GeV). More explicitly 
we show the  $H^{\pm\pm} \rightarrow W^{\pm} W^{\pm}$ dominated branching ratio  by the blueish region,  and  $H^{\pm\pm} \rightarrow l^{\pm} l^{\pm}$ dominated branching ratio  by
yellowish region. The total cross-section has been computed by folding the branching ratios with the cross-section  shown in Fig.~\ref{f:cross HpH0 850}.  In addition, we also include few preliminary cuts, {$ p_{T}({e^{\pm}/\mu^{\pm}})  > 10 \,$ GeV, $ |\eta({e^{\pm}/\mu^{\pm}})| < 2.5 $. For $ M_{H^{0}} = 900\, $GeV and neutrino oscillation parameters to their best fit values\cite{deSalas:2017kay}, we obtain the cut-efficiencies $c_{eff}=0.64$ in $H^{\pm \pm}\rightarrow l^{\pm}l^{\pm}$ mode and $c_{eff} = 0.62$ in $H^{\pm \pm}\rightarrow W^{\pm}W^{\pm}$ mode, that have been included in this analysis.}

\begin{figure}[t]
\begin{center}
\includegraphics[scale=0.18]{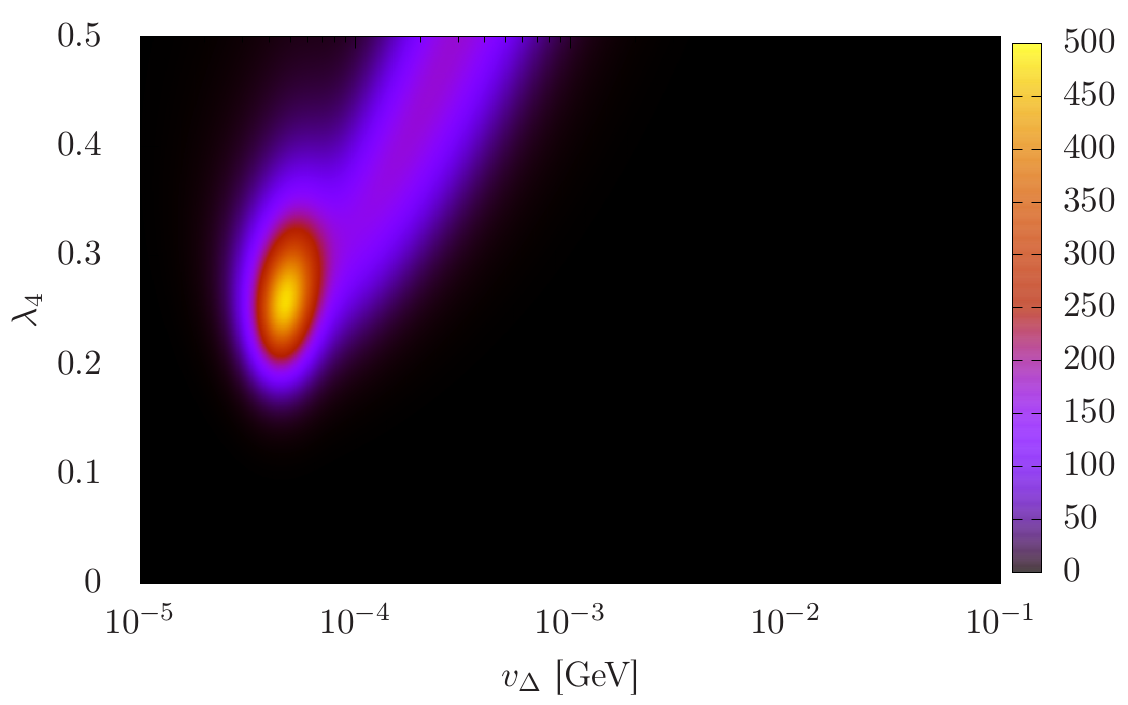}
\includegraphics[scale=0.18]{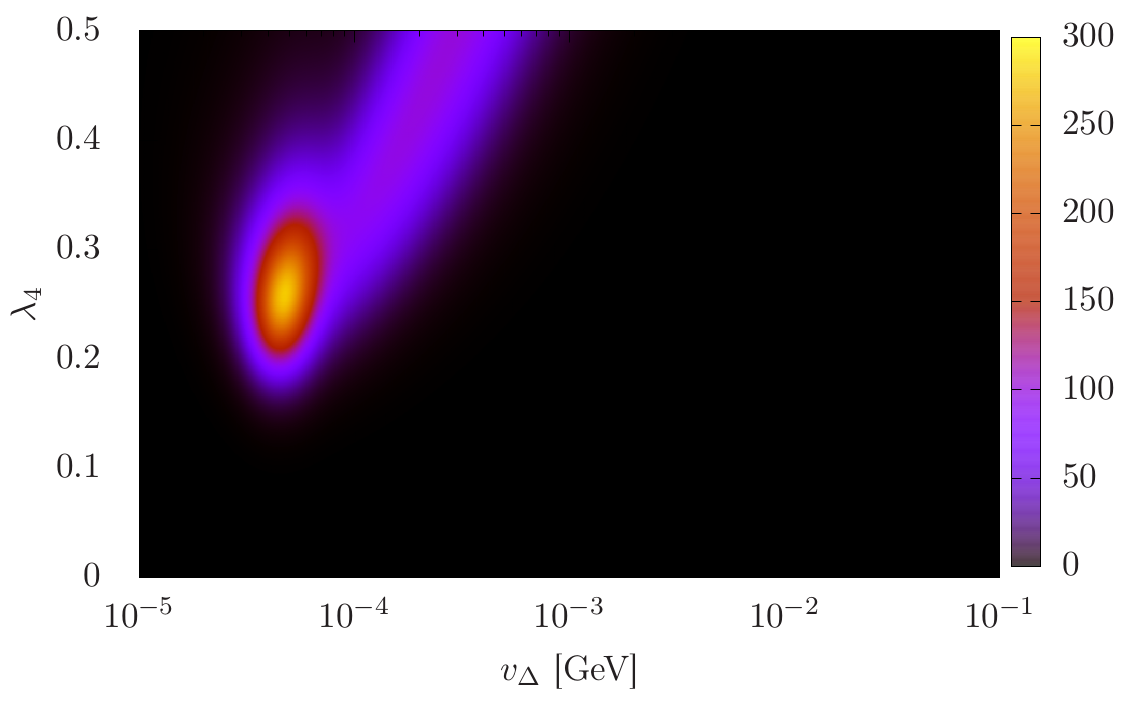}
\includegraphics[scale=0.18]{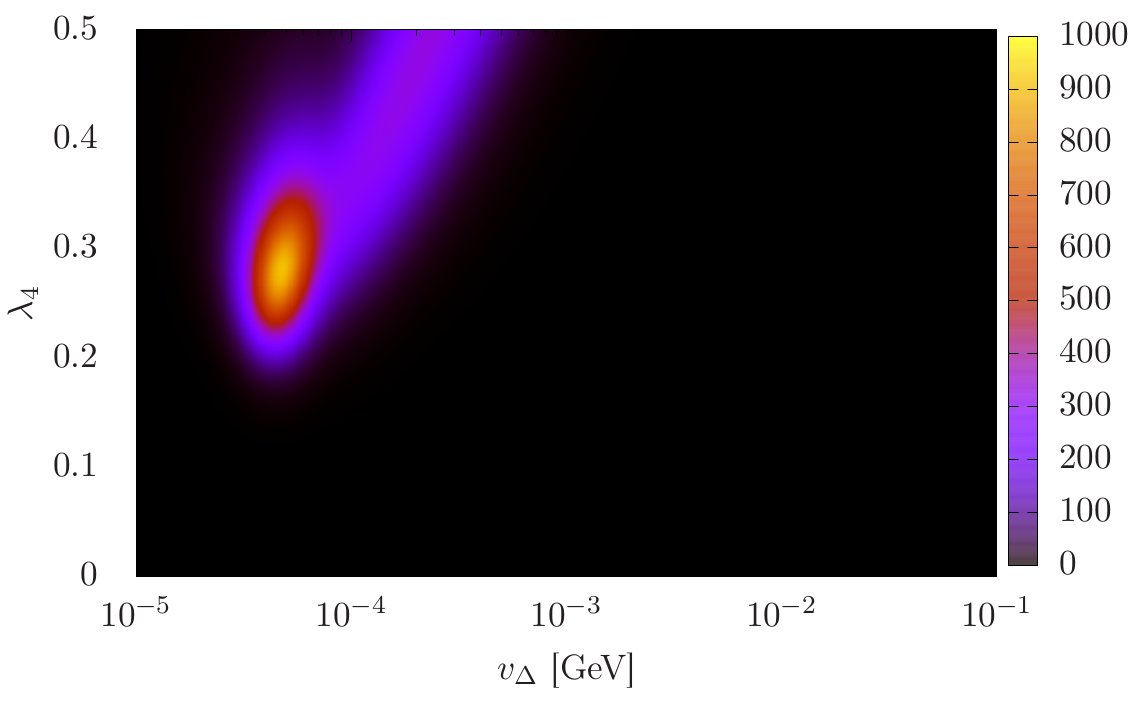}
\caption{These figures represent   number of events for  mass $M_{A^0}/M_{H^0}=900$~GeV. Upper panel: number of same-sign tetralepton events $\ell^+\ell^+\ell^+\ell^+ + X$ from $p p \to H^{+}H^0/A^0$ and subsequent decays (left), number of same-sign tetralepton events $\ell^-\ell^-\ell^-\ell^- + X$ from $p p \to H^{-}H^0/A^0$ and subsequent decays (right). Lower panel: number of same-sign tetralepton events $\ell^+\ell^+\ell^+\ell^+ + Y$ or $\ell^-\ell^-\ell^-\ell^- + Y$ from $p p \to H^{0}A^0$ and subsequent decays. We consider luminosity $\mathcal{L}=30$ $\rm{ab}^{-1}$. For this range of $\lambda_4$, the masses of $H^{\pm}$ and $H^{\pm \pm}$ varies at most by $M_{H^0}-M_{H^{\pm\pm}}\sim$ 8.4 GeV and $M_{H^0}-M_{H^{\pm}}\sim$ 4.2 GeV
respectively.}
\label{event number for mass 850}
\end{center}
\end{figure}

In Fig.~\ref{event number for mass 850}, we  present  the  number  of events for 
 heavier doubly charged Higgs,  charged and neutral Higgs ($\sim 900$ GeV).  
In all the three plots which correspond to $p p \to H^{+}H^0/A^0$,
$p p \to H^{-}H^0/A^0$ and $p p \to H^0A^0$ processes,  its possible to achieve a   significantly large  number of events in a very narrow region indicated
by the yellow patch. This is in contrary to the low mass range, discussed in the previous section,  where we get reasonable number of events for a wider range of triplet vev.   
\begin{figure}[t]
\begin{center}
\includegraphics[scale=0.18]{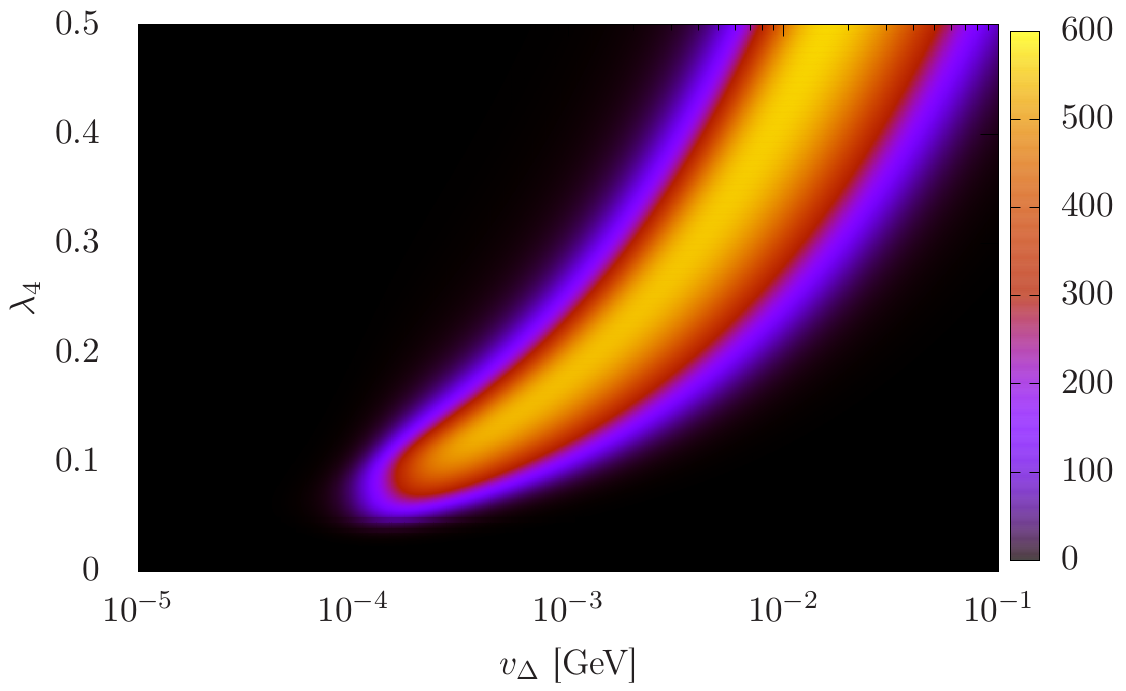}
\includegraphics[scale=0.18]{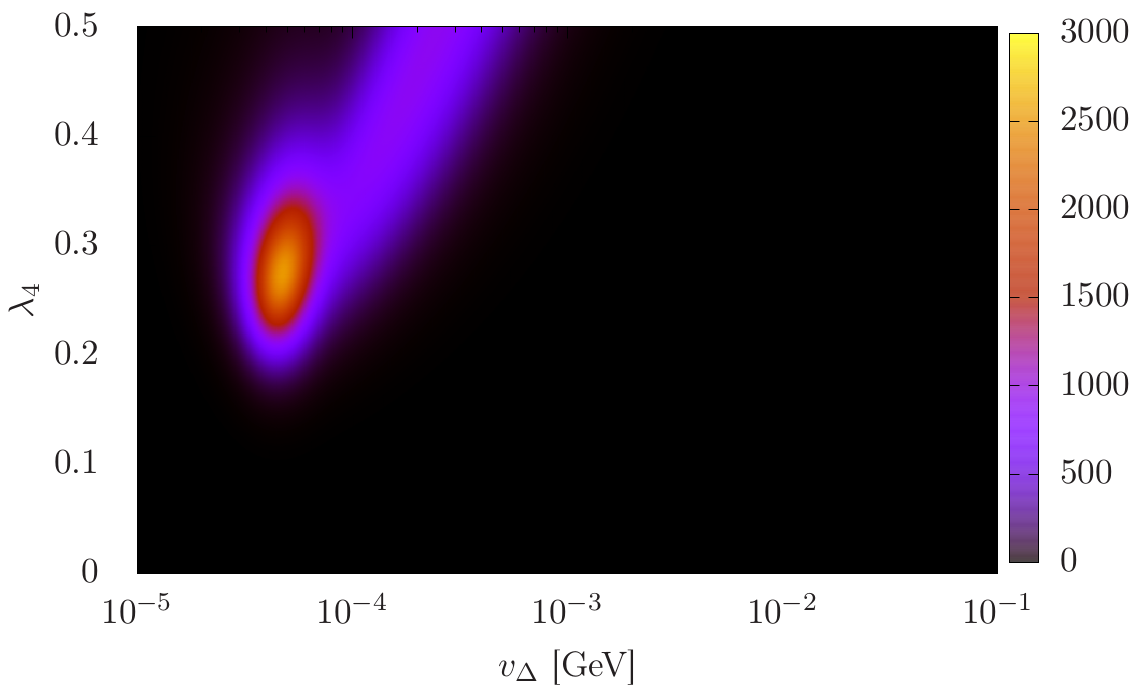}
\caption{Total number of same-sign tetralepton events~($\ell^+\ell^+\ell^+\ell^+ + \ell^-\ell^-\ell^-\ell^-$) for both the cases $M_A^0=253$~GeV~(left figure) and $M_{A^0}=900$~GeV~(right figure).}
\label{total number of events}
\end{center}
\end{figure}
Fig.~\ref{total number of events} represents  the variation of the total number of events for tetra leptons
with either +ve or -ve sign of the leptons. The left panel shows the variation of the sum of the number of events 
($l^{+}l^{+}l^{+}l^{+} + l^{-}l^{-}l^{-}l^{-}$) for the low mass of the particles and its shape is exactly similar
as discussed before (see Fig.~\ref{event number for mass 253}). The figure in the right panel shows the events for 
higher mass and also has the similar shape
as displayed in Fig.~\ref{event number for mass 850}. 

\section{Conclusion \label{conclu}}

The type-II seesaw model is one of the most simplest models of neutrino mass generation, where the model is extended by  an additional triplet scalar field. Due to an extended Higgs sector, and mixing between SM doublet scalar field and triplet scalar, the model contains few additional Higgs fields, including 
doubly charged and singly charged Higgs fields, as well as, CP even and odd neutral Higgs fields.  
In this work, we consider  a type-II seesaw model for neutrino mass generation, and analyse an unique same-sign tetra-lepton signature at $pp$ colliders. This arises  from  the associated production of Higgs fields $H^{\pm} H^0, H^0 A^0$, and the 
subsequent decay of  neutral Higgs field into a singly charged Higgs state, and the decay of a singly charged field into a doubly charged Higgs state. 
More precisely,  for non-degenerate Higgs masses, and for  an intermediate   triplet vev $v_{\Delta}$ in between $10^{-5} \, \textrm{GeV} < v_{\Delta} < 10^{-2}\, \textrm{ GeV}$, the  neutral and charged  Higgs $H^0,A^0$, $H^{\pm}$
 decay predominantly  to $H^0/A^0 \to H^{\pm} W^*$, and $H^{\pm} \to H^{\pm \pm} W^{*}$. The subsequent decays of $H^{\pm \pm}$ either to same-sign di-leptons, or to same-sign gauge bosons lead to the same-sign tetra-lepton final states. 
 We analyse this signature for a $pp$ collider, taking into account two different   c.m.energies, $\sqrt{s}=14$ TeV, and $\sqrt{s}=100$ TeV.  In our analysis, we choose those benchmark mass points, that are consistent with 
 the present limits from 13 TeV LHC.  In particular, for the lower c.m.energy, we explore the tetra-lepton signatures from a lighter Higgs state, and for higher c.m.energy, we consider  a heavier  Higgs states.
 
 As an illustrative example, we first consider a large triplet vev $v_{\Delta} > 10^{-5}$ GeV, and a  benchmark mass with $M_{H^0, A^0}=253$ GeV. We vary the  mass difference between doubly charged Higgs and charge neutral Higgs by at most 5 GeV.  In this region of  triplet vev,  the gauge boson decay mode of $H^{\pm \pm}$ is pre-dominant.  The associated production cross-section $p p \to H^{+} H^0$ varies in between $\sigma \sim 17-20$ fb. The 
 product of different branching ratios  become maximal in a region $v_{\Delta} \sim 10^{-4}\, \textrm{GeV}-10^{-2}$ GeV.  To analyse the signal events, we implement few basic cuts, for which we get a cut efficiency $c_{eff}=0.6$.  With  integrated luminosity of $\mathcal{L}=3000$ $\rm{fb}^{-1}$, 
 we find that a doubly charged Higgs of mass around $257$ GeV can lead to 600 number of events at the future upgrade of LHC. 
 
 Additionally, we also consider  heavier neutral, and doubly charged Higgs, for which  tetra-lepton signature  can be observed  in a $pp$ collider with higher c.m.energy.  For illustration, we consider the mass $M_{H^0, A^0}=900$ GeV,
 and vary the masses of doubly and singly charged Higgs at most by $M_{H^{\pm}}-M_{H^{\pm\pm}}\sim 5$ GeV.   We explore the signal sensitivity for this benchmark point at 100 TeV $pp$ collider.  
 We consider both the  di-lepton and gauge boson decay modes of the doubly charged Higgs. For heavier mass, the branching ratio of $H^{\pm} \to H^{\pm \pm} W^*$ is  large for a very large $\lambda_4$. 
 We find that the production cross-section $p p \to H^{+} H^0$ varies nominally  $\sigma \sim 5$ fb.  
 We find that in a narrow region region in $\lambda_4-v_{\Delta}$ plane, the same-sign tetra-lepton events can be very large $N_{evnt} \sim \mathcal{O}(10^3)$. 
 
\acknowledgments

MM acknowledges the support of the DST-INSPIRE research grant  IFA14-PH-99, and the cluster facility of Institute of Physics (IOP), Bhubaneswar, India. EJC and MM thank the workshop 'NuHoRizon', held during March at  HRI, India, where the work has been initiated.    
\bibliographystyle{JHEP}
\bibliography{t2seesaw.bib}

\end{document}